\documentclass[fleqn,usenatbib]{mnras}

\usepackage{newtxtext,newtxmath}

\usepackage[T1]{fontenc}

\DeclareRobustCommand{\VAN}[3]{#2}
\let\VANthebibliography\thebibliography
\def\thebibliography{\DeclareRobustCommand{\VAN}[3]{##3}\VANthebibliography}

\usepackage{graphicx}	
\usepackage{amsmath}	
\usepackage{url}
\usepackage{stfloats} 
\usepackage{ulem}
\usepackage{bm}
\usepackage{caption}

\title[Axe-shaped radio galaxy]{Investigation of the Axe-shaped Radio Galaxy J1051+5523 with uGMRT}

\author[Sudheesh et al.]{
Sudheesh T. P.,$^{1}$\thanks{E-mail: sudheesh.tp@res.christuniversity.in}
Ruta Kale,$^{2}$
V. Jithesh,$^{1}$
Ramananda Santra,$^{2}$
C. H. Ishwara-Chandra$^{2}$
and Joe Jacob$^{3}$
\\
$^{1}$Department of Physics and Electronics, Christ University, Hosur Main Road, Bengaluru 560029, India.\\
$^{2}$National Centre for Radio Astrophysics, TIFR, Post Bag No. 3, Ganeshkhind, Pune 411007, India\\
$^{3}$Newman College, Thodupuzha, Kerala 685584, India
}

\date{Accepted XXX. Received YYY; in original form ZZZ}

\pubyear{\the\year{}}

\begin{document}
\label{firstpage}
\pagerange{\pageref{firstpage}--\pageref{lastpage}}
\maketitle

\begin{abstract}
We present a multi-frequency study of the bent-tail radio galaxy J1051+5523, located in the galaxy cluster WHL J105147.4+552309. This wide-angle tail (WAT) galaxy exhibits a complex radio morphology, characterised by a right-angled bend in the northern jet, which resembles an axe, and multiple kinks in the southern jet, as observed in the deep uGMRT radio observations. The radio power of J1051+5523 at 150 MHz is estimated to be $2.91 \times 10^{25}\,\mathrm{W\,Hz^{-1}}$, placing it in the transition zone between FRI and FRII radio galaxies. The spectral index map reveals a flat core and relatively flat lobes, which may indicate ongoing particle acceleration or a relatively young population of relativistic electrons in the lobes. Further, we estimate the equipartition magnetic fields, and spectral ages of the northern and southern lobes to be approximately 150 Myr and 153 Myr, respectively, suggesting a long-lived radio source with sustained AGN activity. A relative velocity of 278 $\pm$ 2643 $\mathrm{km\,s^{-1}}$ is obtained for the host galaxy. Due to the large uncertainty associated with the relative velocity estimates, the contribution of ram pressure to the jet bending remains inconclusive. The low mass of the host cluster ($\sim 2 \times 10^{14}\,M_\odot$) and the lack of diffuse X-ray emission indicate a reduced likelihood of major mergers, but minor mergers or interactions remain possible. We propose that the observed WAT morphology of J1051+5523 is likely shaped by a combination of ram pressure and/or buoyant forces within the cluster environment.
\end{abstract}

\begin{keywords}
radio continuum: galaxies-galaxies: clusters: intracluster medium-galaxies: active-radiation mechanisms: non-thermal
\end{keywords}



\section{Introduction}

The radio galaxy phase is considered a significant stage in the life cycle of massive galaxies, influenced by a variety of factors. The main emission mechanism is synchrotron radiation \citep{1956burbidge}. The activity of the supermassive black hole (SMBH) at the galaxy's core drives radio galaxies, forming twin, collimated, and relativistic jets \citep{1974blandford, 2020hardcastle}. When these jets are sufficiently powerful, they can expand into a cocoon \citep[e.g.][]{1974scheuer, 1991falle}, which can extend into the host galaxy's interstellar medium (ISM) and later the intergalactic medium (IGM). According to the radio morphological classification system introduced by \citet{1974fanaroff}, radio galaxies are categorised into two primary classes: FR I and FR II. This classification is based on the ratio of the distance between the regions of highest radio surface brightness on opposite sides of the central nucleus to the total extent of the source. FR I galaxies are identified as edge-darkened radio sources, while FR II galaxies are identified as edge-brightened sources.

In addition to the basic classifications mentioned above, radio galaxies exhibit a variety of morphologies. Some of the predominant morphologies include wide-angle tailed radio galaxies (WATs) \citep{1976owen}, narrow-angle tailed radio galaxies (NATs) \citep{1968ryle, 1972miley}, X-shaped radio galaxies (XRGs) \citep[e.g.][]{1984leahy}, and Z/S-shaped radio galaxies (ZRGs) \citep[e.g.][]{1972riley}. These morphologies are collectively identified under the common name bent-tail (BT) radio galaxies. It is widely accepted that the bending of the radio jets in these galaxies is likely caused by the ram pressure from the environment in which they are residing \citep{1979jones, 1980burns, 2008freeland}.

WAT galaxies are regarded as an intermediate class of radio galaxies, positioned between FR I and FR II types \citep{2019missaglia}. They are described as having well-collimated jets that extend for several kiloparsecs before suddenly spreading into diffuse and frequently bent plumes \citep{2019missaglia}. WAT radio galaxies are generally associated with a dominant galaxy in the galaxy cluster \citep[e.g.][]{1976owen, 1986burns, 1993odonoghue}. \citet{2023odea}, in their recent review, suggests that WATs serve as indicators of merging galaxy clusters, as they are typically hosted by the brightest cluster galaxies (BCGs) in such clusters. However, WAT galaxies are also observed in galaxy groups, filaments, and even in field environments \citep[e.g.][]{2010edwards, 2018obrien, 2019garon, 2022morris}. Various theories have been suggested in the literature to explain the phenomenon of tail bending in various bent-tail radio galaxies \citep[e.g.][]{1982burns, 1985bodo, 1994venkatesan, 1995worrall}. 

In the case of WATs, \citet{2004hardcastle} suggested different processes through which their jets terminate, before expanding into plumes. Further, depending on the length of the jets before they bend, various mechanisms of bending have been proposed by different authors \citep[e.g.][]{1986burns, 1993odonoghue}. According to \citet{1986burns}, the smallest WATs may bend due to dynamic pressure resulting from the movement of the radio galaxy around another large galaxy or a subcondensation of galaxies within the cluster. Additionally, moderate-sized, symmetrically bent WATs may show bending as a jet passes through a steep pressure gradient. In contrast, large WATs associated with BCGs exhibit sharp and irregular bends in their jets and are likely to interact with clouds in the cluster's gas. In overview, the physical mechanisms that initiate bending in WAT radio sources are complex and remain under investigation.

This paper discusses the properties of an archetypal wide-angle tail (WAT) radio galaxy J1051+5523 (RA = 10:51:47.39, Dec = +55:23:08.39 \citep{2020ahumada}). It is also identified in NASA Extragalactic Database (NED) as NVSS J105147+552316. We were conducting a pilot study to identify WAT radio sources from the FRI catalogue by \citet{2017capetti}. Using various cluster catalogues, we examined the cluster associations of these WAT sources and shortlisted those associated with clusters where the host galaxy is also the Brightest Cluster Galaxy (BCG). During a visual inspection of the radio morphologies of the shortlisted sources in various all-sky radio surveys, we identified this particular radio source. The TIFR-GMRT Sky Survey \citep[TGSS; ][]{2017intema} image of J1051+5523 shows a WAT morphology in which the northern jet exhibits a near right-angled bend, whereas the southern jet lacks such a feature. This asymmetry in jet bending and its potential causes have not been explained for J1051+5523 in the literature. \citet{2017capetti} classified the radio galaxy as an FR I source, while \citet{2011koziel} classified it as an FR II source. However, \citet{2011koziel} could not identify any hotspots, despite classifying it as an FR II source. \citet{2011wing} visually classified it as a normal bipolar FR I radio galaxy. \citet{2017miraghaei} also classified the galaxy as a normal FR I galaxy and did not identify it as a WAT source. However, they classified the host galaxy as a High-Excitation Radio Galaxy (HERG). They also estimated the host galaxy parameters and the environmental parameters of J1051+5523. The first identification of J1051+5523 as a bent-tail source is found in the studies of \citet{2019mazhixian}. They reported it as an FR II-like bent-tail source, using a convolutional neural network-based autoencoder. J1051+5523 is located within the galaxy cluster WHL J105147.4+552309 \citep{2012wen}. It is hosted by the optical galaxy SDSS J105147.39+552308.3, at a redshift of 0.07393 $\pm$ 0.00002 \citep{2020ahumada}, which is also identified as the Brightest Cluster Galaxy (BCG) of the cluster \citep{2012wen}. \citet{2015zhao} reported the host galaxy as a cD galaxy also.  

Even though it is not identified as a WAT source, \citet{2011wing} observed that the galaxy resides in a relatively dense environment with a richness of 62. The authors defined richness as the number of galaxies within 1 Mpc of the radio source with an absolute r-magnitude brighter than $\mathrm M_{r}$ = -19, corrected for background counts. \citet{2012wen} estimated the cluster's richness to be 34.45, with 27 galaxy member candidates within the cluster’s $\mathrm r_{200}$ radius. \citet{2017tempel} also studied the clustering using a modified friends-of-friends algorithm. The authors estimated a richness of 35 for the cluster. They also checked for the possibility of cluster mergers but found no evidence of such events associated with this cluster. However, a deep, multi-frequency radio study of the morphology of J1051+5523, and its potential causes have not yet been explored in detail. In this paper, we present deep, multi-frequency radio observations of the archetypal WAT radio source, using uGMRT. 

This paper is organized as follows. In Sect.~\ref{sec:observations} we describe the uGMRT observations and data analysis of J1051+5523. The radio properties of the source and the results are then presented in Sect.~\ref{sec:results}. A detailed discussion is presented in Sect.~\ref{sec:discussions}. Section~\ref{sec:conclusion} explains our summary and conclusions.

In this study, we utilized a flat $\Lambda$CDM cosmological model based on the findings from the Planck Collaboration ($H_{0} = 67.8~\mathrm{km\,s^{-1}\,Mpc^{-1}}$, $\Omega_{m}$ = 0.308 and $\Omega_{\Lambda}$ = 0.692; \citep{2016planck}). At $z$ = 0.07393, this corresponds to a scale of 1.450 kpc/". We define spectral index $\alpha$ such that $S_{\nu} \propto \nu^{\alpha}$, where S$_{\nu}$ is the flux density at a given frequency $\nu$.

\section{Observations and Data Analysis}
\label{sec:observations}

We observed J1051+5523 (Project Code: 41\_100, PI: Sudheesh T P) using the upgraded Giant Metrewave Radio Telescope (uGMRT), which features a near-continuous frequency coverage from 120 to 1450 MHz \citep{2017gupta}. Our observations were carried out at Bands 3 and 4 using the GMRT Wideband Backend \citep[GWB;][]{2017reddy}. The Band 4 observation was done on 3 November 2021 and the Band 3 observation was done on 27 November 2021. We have summarised the observation details in Table~\ref{tab:ugmrt-obs-summary}.

 \begin{table}
	\centering
	\caption{Summary of uGMRT observations}
	\label{tab:ugmrt-obs-summary}
	\begin{tabular}{lcc} 
		\hline
		   & Band 3 & Band 4 \\
		\hline
		Frequency range (MHz) & 300-500 & 550-750\\
		Number of channels & 2048 & 2048\\
		Bandwidth (MHz) & 200 & 200\\
            On source time (hr) & 3 & 3\\
            Integration time (sec) & 8 & 8\\
            Flux calibrator & 3C147, 3C286 & 3C147, 3C286\\
            Phase calibrator & 0834+555 & 0834+555\\
		\hline
	\end{tabular}
\end{table}

\par We reduced the uGMRT data using the {\tt CAPTURE}\footnote{\url{https://github.com/ruta-k/CAPTURE-CASA6}} pipeline \citep{2021kale}. {\tt CAPTURE} is a continuum imaging pipeline which uses the Common Astronomy Software Applications \citep[{\tt CASA}; ][]{2007mcmullin} to process the uGMRT data. Initially, the data were flagged for bad antennas and Radio Frequency Interference (RFI), and then the flux density of the primary calibrators was set using the Perley–Butler 2017 flux scale \citep{2017perley}. Following standard calibration procedures, the data of the target source were calibrated. The calibrated data was then split, and further flagged using different flagging modes. To manage data volume while avoiding bandwidth smearing, 10 frequency channels were averaged, resulting in a channel width of around 1 MHz per channel. These processes were applied consistently to both bands 3 and 4. The target visibilities were then imaged with the {\tt CASA} task {\tt tclean}. During imaging and self-calibration, the target source measurement set was divided into 8 frequency sub-bands, each containing 20 channels. This method splits the total 200 MHz bandwidth into 8 spectral windows, performing imaging and self-calibration on the sub-banded data file. To perform the primary beam correction of the images, we used the task {\tt ugmrtpb}\footnote{\url{https://github.com/ruta-k/uGMRTprimarybeam}}.

\section{Results}
\label{sec:results}
\begin{figure}
	\includegraphics[width=\columnwidth]{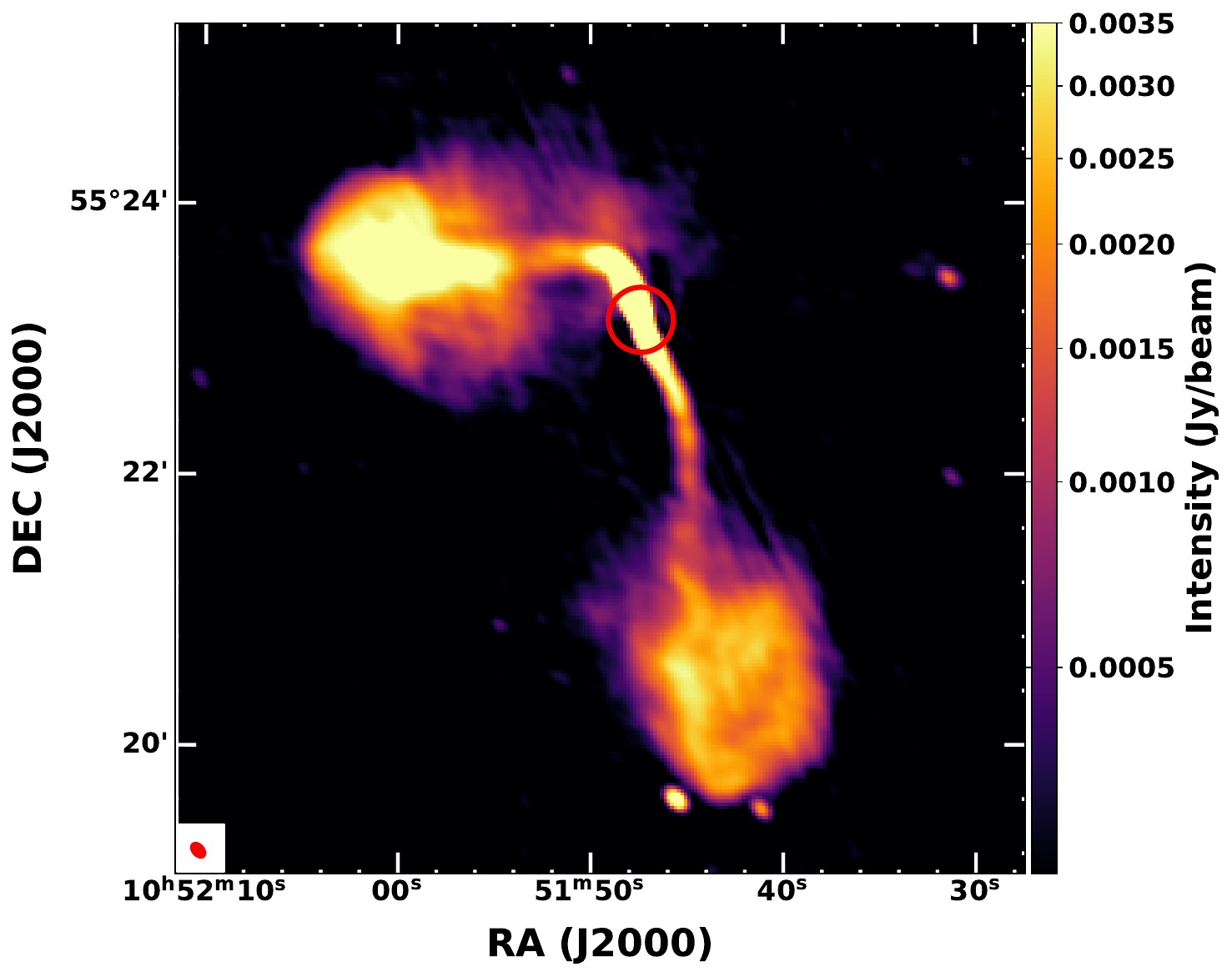}
 	\includegraphics[width=\columnwidth]{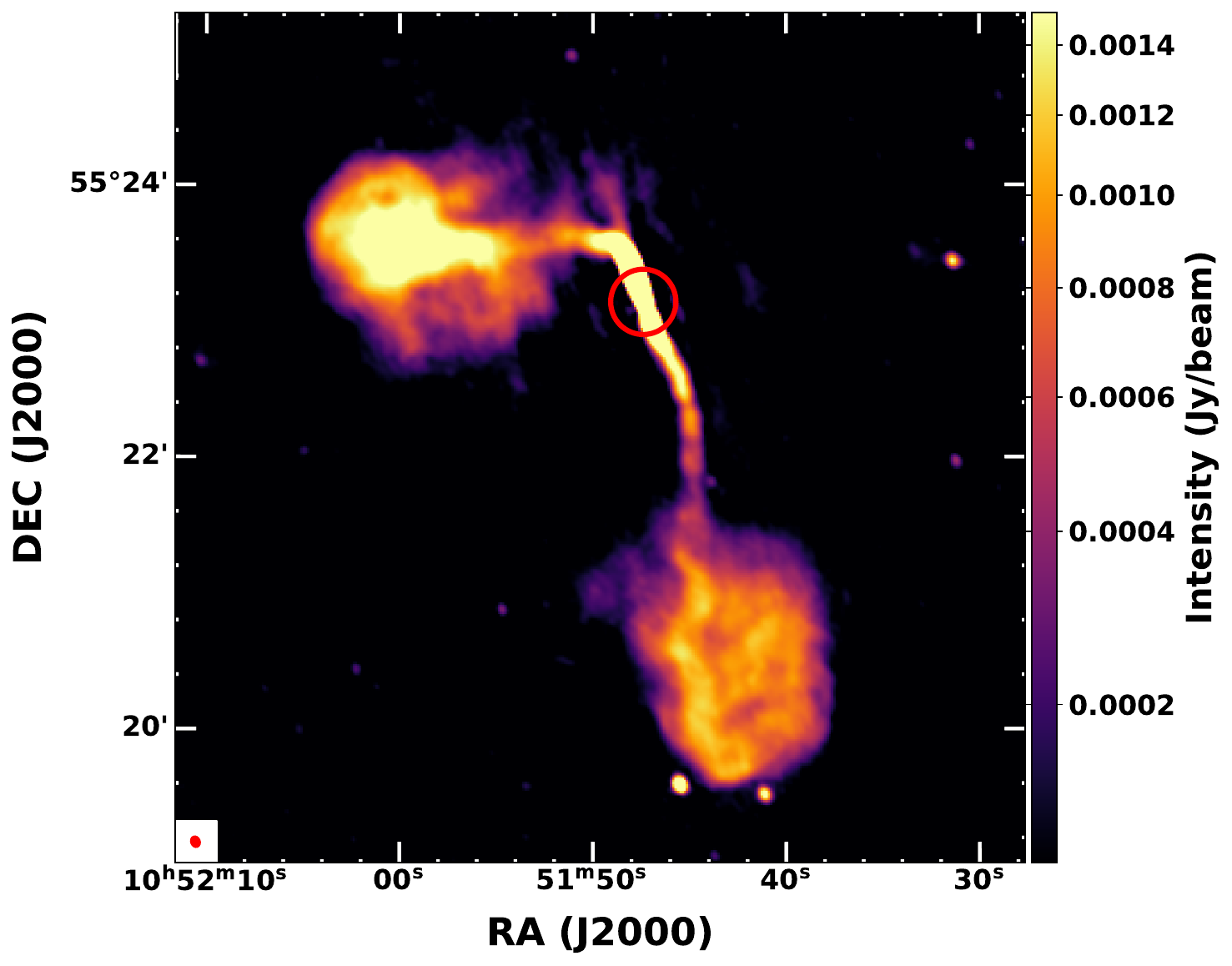}
    \caption{Top panel: uGMRT Band 3 image of J1051+5523. Bottom panel: uGMRT Band 4 image of J1051+5523. The red circle shows the position of the host galaxy. The radio beam is shown at the bottom left corner of each image.}
    \label{fig:b3-b4-image-J1051+5523}
\end{figure}

\begin{figure}
        \centering
	\includegraphics[width=\columnwidth]{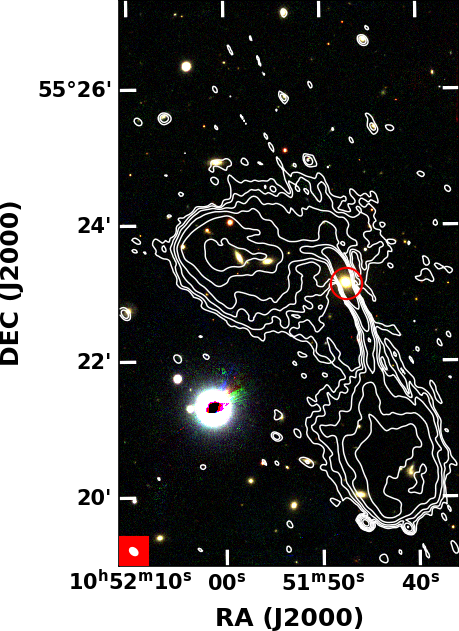}
    \caption{Pan-STARRS colour composite image of SDSS J105147.39+552308.3. The image is an RGB composition of Pan-STARRS g (blue), r (green), and i (red) band images. The red circle shows the position of the host galaxy. The uGMRT Band 3 contours are overlaid over the optical image with six levels in multiples of 2, starting from 3$\sigma$ where $\sigma$ = 0.044 mJy/beam. The bottom left corner shows the radio beam.}
    \label{fig:rgb-image}
\end{figure}

The uGMRT Band 3 and Band 4 images of J1051+5523 are given in Fig.~\ref{fig:b3-b4-image-J1051+5523}. The basic details of the images are given in Table~\ref{tab:b3-b4-image_details}. J1051+5523 is hosted by the optical galaxy SDSS J105147.39+552308.3, which is at a redshift of 0.07393 $\pm$ 0.00002 \citep{2020ahumada}. The host galaxy has been identified as the Brightest Cluster Galaxy (BCG) in the galaxy cluster WHL J105147.4+552309 \citep{2012wen}. A Pan-STARRS RGB colour composite image of SDSS J105147.39+552308.3 with Band 3 radio contours overlaid is provided in Fig.~\ref{fig:rgb-image}.

\begin{table}
	\centering
	\caption{Details of the Band 3 and Band 4 images of J1051+5523 shown in Figure~\ref{fig:b3-b4-image-J1051+5523}.}
	\label{tab:b3-b4-image_details}
	\begin{tabular}{lccc}
		\hline
		Band & rms & Resolution & PA\\
             & (mJy/beam) & (arcsec $\times$ arcsec) & (degree)\\
		\hline
		Band 3 & 0.044 & 7.7" $\times$ 5.2" & 41.5\\
		Band 4 & 0.021 & 4.9" $\times$ 3.9" & 21.2\\
		\hline
	\end{tabular}
\end{table}

\subsection{Linear Size}
Our deep observations of J1051+5523 at Bands 3 and 4 of uGMRT show diffuse emission on a large scale. Both Band 3 and Band 4 should detect the entire emission, given the source's size and the largest linear size recovered by the telescope at these frequencies. For Band 3, it is 32 arcmin, and for Band 4, it is 16 arcmin. However, as Band 3 corresponds to a lower frequency, it likely has better sensitivity to extended regions of low surface brightness, possibly providing a more accurate estimate of the Largest Angular Size (LAS). Thus, we used the Band 3 radio map to estimate the LAS of the radio source. The lowest contour level at 3\,$\sigma$ is estimated to be 0.132 mJy/beam. The LAS for the emission above 3\,$\sigma$ is then estimated to be 5.6$^\prime$. At the redshift of $z$ = 0.07393, this corresponds to a projected linear size of 487.6 kpc.

\subsection{Flux Density and Radio Power}
\label{sec:flux-density}
The integrated flux densities of J1051+5523 at Bands 3 and 4 were measured using the {\tt imview} task of the {\tt CASA}. To estimate the flux density, we manually selected the regions of emission above the 3\,$\sigma$ level in the Band 3 and Band 4 radio maps. The same region was used to estimate the flux densities from both Band 3 and Band 4 radio maps. The region used is shown in Fig.~\ref{fig:region-for-flux-density}. There are two point sources seen close to the southern lobe. They were not related to our target source. Their flux densities were measured from the Band 3 and Band 4 radio maps using the {\tt CASA} task {\tt imfit}, and subtracted from the integrated flux densities. The uncertainty in flux density is calculated using the scheme provided by \citet{2003klein} as:

\begin{equation}
    \delta S = \sqrt{\left(S_{\nu}\times \delta S_{\mathrm{c}}\right)^2+\left(\sigma\times \sqrt{\frac{\Omega_{\mathrm{int}}}{\Omega_{\mathrm{beam}}}}\right)^2}
    \label{eq:fluxerr}
\end{equation}

\noindent where $\delta S$ is the uncertainty in flux density, $S_{\nu}$ is the flux density measured in mJy at frequency $\nu$, $\delta S_{\mathrm{c}}$ is the calibration error, $\sigma$ is the rms noise of the radio map, $\Omega_{\mathrm{int}}$ is the size of the integration area and $\Omega_\mathrm{{beam}}$ is the beam solid angle. For Band 3 and Band 4 data, the calibration error $\delta S_{\mathrm{c}}$ is taken as 10\% \citep{2017chandra, 2024santra}. The flux density of J1051+5523 in Band 3 is estimated as 1185 $\pm$ 118.5 mJy and in Band 4 is estimated as 945 $\pm$ 94.5 mJy. 

The radio power of J1051+5523 was estimated using the relation given in \citet{2009donoso} as:

\begin{equation}
    P_{\nu} = 4\pi D_{\mathrm{L}}^2S_{\nu}(1+z)^{-(1+\alpha)}
    \label{eq:radpower}
\end{equation}

\noindent where $P_{\nu}$ is the radio power, $D_{\mathrm{L}}$ is the luminosity distance at the redshift $z$, $S_{\nu}$ is the flux density and $(1+z)^{-(1+\alpha)}$ is the standard k correction term where $\alpha$ is the radio spectral index. The radio power of J1051+5523 at Band 3 is computed as $1.65 \times 10^{25}\,\mathrm{W\,Hz^{-1}}$, and at Band 4 is $1.31 \times 10^{25}\,\mathrm{W\,Hz^{-1}}$.

\subsection{Spectral Index}
\label{sec:specindex}
In addition to our uGMRT Band 3 and Band 4 data, J1051+5523 has been observed in several other sky surveys. Specifically, the source has been detected in the LOFAR Two-Metre Sky Survey \citep[LoTSS; ][]{2019shimwell, 2022shimwell}, the TIFR-GMRT Sky Survey \citep[TGSS; ][]{2017intema}, and the NRAO VLA Sky Survey \citep[NVSS; ][]{1998condon}. The LoTSS radio map has a local rms noise of 0.11 mJy/beam, significantly lower than the 2.12 mJy/beam measured in the TGSS map. The signal-to-noise ratio (SNR) histogram of TGSS also shows a broader distribution, indicating a higher noise scenario. However, the LoTSS image contains residual phase-error-like artefacts, likely due to calibration imperfections. This led us to prefer the TGSS data over the LoTSS data in the 150 MHz regime. The radio maps of J1051+5523 from TGSS and NVSS were retrieved from their respective archives, and the corresponding flux densities were measured as detailed in Section~\ref{sec:flux-density}. We used the same region mentioned in Section~\ref{sec:flux-density} to extract the flux density from both TGSS and NVSS radio maps. However, the region encloses two point sources near the southern lobe, which are not resolved in both the radio maps. Using Band 3 and Band 4 flux densities of the point sources, we estimated the two-point spectral index of the two sources and used it to arrive at an approximate estimate of their flux densities at TGSS and NVSS frequencies. The estimated flux densities of the point sources in TGSS and NVSS frequencies are then subtracted from the respective integrated flux densities measured. The TGSS and NVSS radio maps, together with the region used for extracting flux density, overlaid are shown in Fig.~\ref{fig:nvss-tgss-radiomaps}. Furthermore, the source was also observed using the NRAO Green Bank Telescope (GBT) by \citet{1991becker}. We obtained the flux density of J1051+5523 using GBT, which is provided in \citet{1991becker}. The summary of the observed spectrum of J1051+5523 is provided in Table~\ref{tab:observed-spectrum}.

\begin{figure}
	\includegraphics[width=\columnwidth]{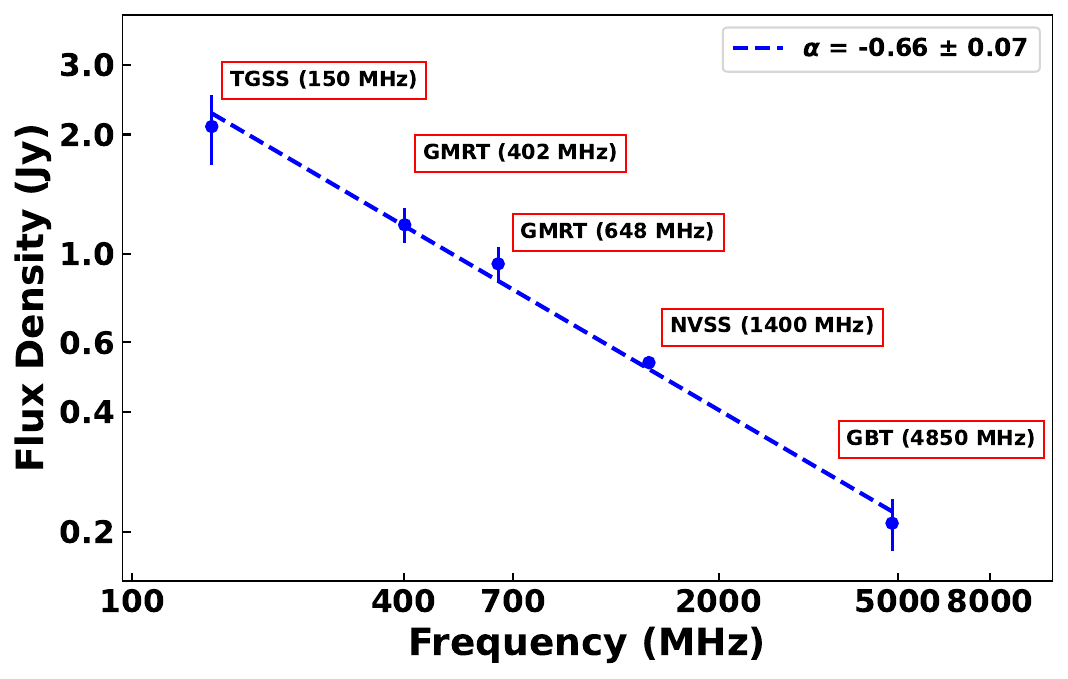}
    \caption{The integrated spectrum of J1051+5523. The flux densities at different frequencies are listed in Table~\ref{tab:observed-spectrum}. The error bars indicate the uncertainties in the respective flux densities, which are estimated using Equation~\ref{eq:fluxerr}.}
    \label{fig:IC644-integrated-spectrum}
\end{figure}

\begin{table*}[h]
        \caption{Observed spectrum of J1051+5523. The flux density from GBT is obtained from \citet{1991becker}. The references for calibration errors are: TGSS \citep{2020dabhade}, uGMRT Bands 3 and 4 \citep{2017chandra}, NVSS \citep{1998condon}, and GBT \citep{1991becker}.}
	\label{tab:observed-spectrum}
        \begin{tabular}{lcccc}
		\hline
		Survey/ & Frequency & Calibration Error & Flux Density & rms\\
           Telescope & (MHz) & $\%$ & (Jy) & ($\mu$Jy/beam) \\
		\hline
            TGSS & 150 & 20 & 2.095 $\pm$ 0.42 & 2120\\
            uGMRT Band 3 & 402 & 10 & 1.185 $\pm$ 0.12 & 44\\
            uGMRT Band 4 & 648 & 10 & 0.945 $\pm$ 0.09 & 21\\
            NVSS & 1400 & 3 & 0.533 $\pm$ 0.02 & 510\\
            GBT & 4850 & 15 & 0.21 $\pm$ 0.03 & 6500\\
		\hline
	\end{tabular}
\end{table*}
The integrated spectrum of J1051+5523 is given in Fig.~\ref{fig:IC644-integrated-spectrum}. The observed spectrum is accurately fitted by a single power law within the frequency range of 150-4850 MHz, yielding an integrated radio spectral index of -0.66 $\pm$ 0.07.

\subsection{Spectral Index Map}
\label{sec:specmap}
\begin{figure*}
    \includegraphics[width=7in]{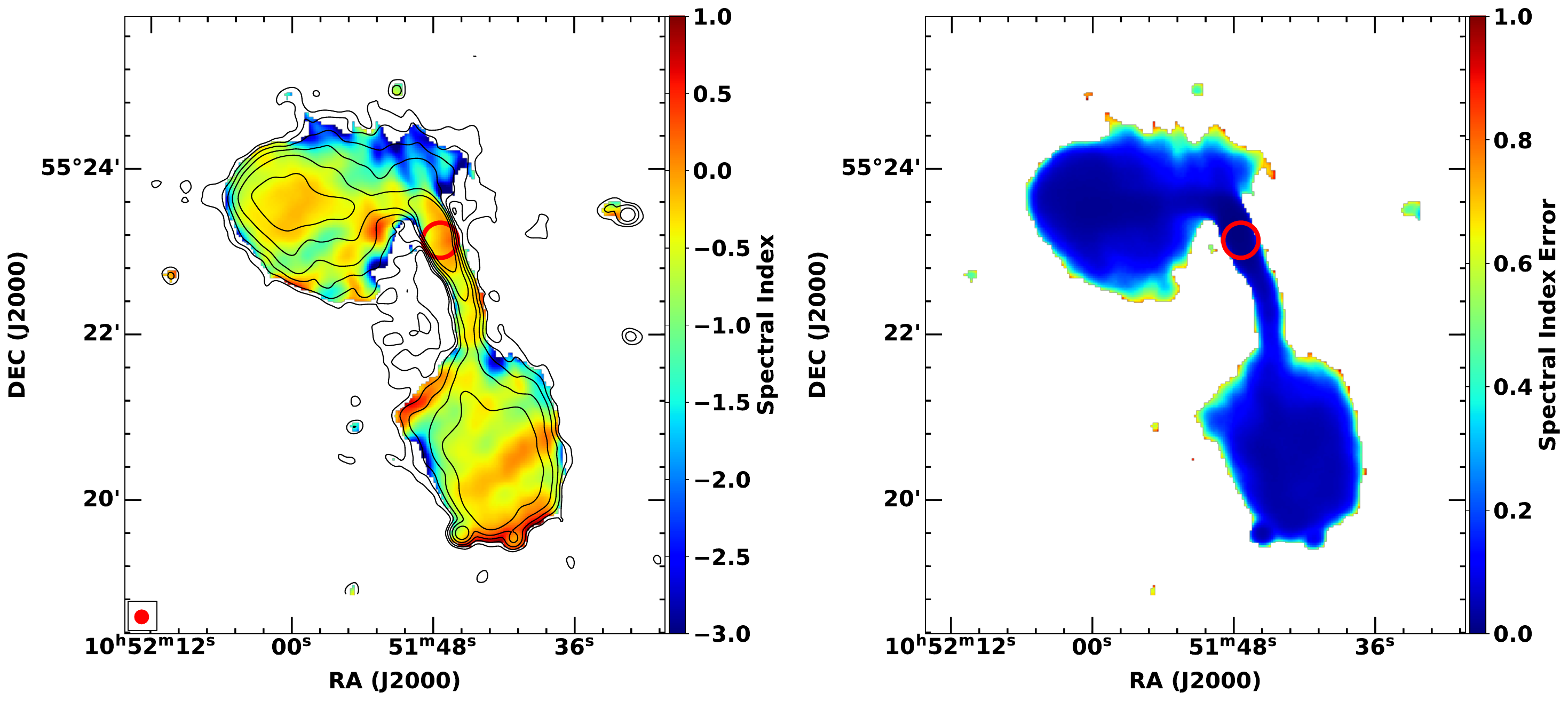}
    \caption{Left panel: Spectral index map of J1051+5523 between Band 3 (402 MHz) and Band 4 (648 MHz). The spectral index map is overlaid with Band 3 radio contours in black, drawn at $3 \sigma \times$[1, 2, 4, 8, 16, 32], where $\sigma$ = 0.083 mJy/beam, is the local rms noise, derived from the Band 3 radio map mentioned in Section~\ref{sec:specmap}. The red circle shows the position of the host galaxy. Radio beam is shown at the bottom left corner. Right panel: Spectral index error map.}
    \label{fig:b3-b4-specindexmap}
\end{figure*}

Spatially resolved spectral index maps are highly effective in elucidating the various physical processes responsible for the bent morphology observed in J1051+5523. Using uGMRT Band 3, and Band 4 data, we plotted the spectral index map of J1051+5523. Band 3 and 4 radio maps are re-imaged with the same UV range of $0.16~\mathrm{k}\lambda \;-\; 38~\mathrm{k}\lambda$ with uniform weighting scheme to match the spatial scale of the radio emission at both frequencies. The radio maps are then convolved to a common resolution of 10\arcsec\ $\times$ 10\arcsec. For each radio map, only those pixels with flux values greater than $3\sigma$ are taken, where $\sigma$ is the local rms around the source. However, instead of taking the absolute flux density value from each pixel, a Gaussian flux density distribution is defined with the flux density of the pixel as the mean flux density, and the standard deviation as the background noise ($\sigma$) of the respective radio map. From this distribution, 1000 flux density values are randomly drawn, whose mean will give the flux density of the corresponding pixel. The uncertainty in the flux density of the pixel is taken as the standard deviation of the 1000 extracted values. The method is used in \citet{2017degasperin} and \citet{2024santra}. The Band 3-Band 4 spectral index map of J1051+5523 is given in the left panel of Fig.~\ref{fig:b3-b4-specindexmap}, and the corresponding error map is shown in the right panel.

The spectral index map shows a flat core and two lobes with an average spectral index ($\alpha$) $\sim$ -0.5. The northern lobe shows some regions with steep spectral index ($\alpha\ <$ -1.0). The lobe also shows two regions of relatively flat spectral index ($\alpha\ >$ -0.5) separated by a region of relatively less flat spectral index ($\alpha\ <$ -0.5). Compared to the northern lobe, the southern lobe has a relatively uniform spectral index with $\alpha \sim$ -0.5. The southern lobe shows some spectral index structure. They may indicate spectral gradients due to backflow of synchrotron plasma or localised particle re-acceleration, although projection effects and resolution may also contribute. In line with standard synchrotron ageing models, such gradients may arise as particles are carried away from the acceleration sites.

\subsection{Magnetic Field and Spectral Age}
Assuming a minimum energy condition, we have determined the minimum energy density and magnetic field for J1051+5523. The magnetic field is estimated under minimum energy conditions by minimising the total energy. However, this estimate of the magnetic field is approximately equivalent to the estimates using the equipartition assumption \citep{1980miley}. Thus, we refer to our magnetic field as the equipartition magnetic field. The minimum energy density was derived using the formula presented by \citet{2004govoni} as:

\begin{equation}
    u_\mathrm{{min}} = \xi (\alpha,\nu_\mathrm{{1}},\nu_\mathrm{{2}})(1+k)^{4/7}(\nu_\mathrm{{0}})^{4\alpha /7}(1+z)^{(12+4\alpha) /7}(I_\mathrm{{0}}/d)^{4/7}
    \label{eq:minenergy}
\end{equation}

This derivation is based on the assumption of a uniform magnetic field and an isotropic particle distribution. Here, $u_\mathrm{{min}}$ represents the minimum energy density, $k$ denotes the ratio of the energy of relativistic protons to that of electrons, and $\nu_\mathrm{{1}}$ and $\nu_\mathrm{{2}}$ are the lower and upper frequency limits within which the spectrum is integrated. $I_\mathrm{{0}}$ refers to the surface brightness in $\mathrm{mJy}\,\mathrm{arcsec}^{-2}$ measured at a frequency of $\nu_\mathrm{{0}}$ MHz. The parameter $\xi (\alpha,\nu_\mathrm{{1}},\nu_\mathrm{{2}})$ is calculated as described by \citet{2004govoni} for different spectral index values. $d$ indicates the depth of the source in kpc. For this calculation, we have assumed $\nu_\mathrm{{1}}$ = 10 MHz, $\nu_\mathrm{{2}}$ = 10 GHz, and $k$ = 1. We utilised the lowest frequency TGSS 150 MHz radio map for our estimations. Assuming a cylindrical shape for the radio source, we calculated the source depth $d$ to be 145.42 kpc. The surface brightness $I_\mathrm{{0}}$ was also measured from the TGSS radio map at a frequency of $\nu_\mathrm{{0}}$ = 150 MHz. From the integrated spectrum, the spectral index $\alpha$ was determined, and we assumed a filling factor of unity.

The magnetic field of J1051+5523 is then calculated using the relation:

\begin{equation}
    B_{\mathrm{eq}} = \left(\frac{24\pi}{7} u_\mathrm{{min}}\right)^{1/2}
    \label{eq:magfield-eq}
\end{equation}

Equation~\ref{eq:magfield-eq} is based on classical formalism. It has a few limitations. The estimation is based on the hardly known parameter $k$. Further, this formalism traces the electron energy spectrum only within a limited range of frequencies. Consequently, the estimated total energy ratio may not accurately reflect the actual energy distribution. To address these limitations, we have also employed the revised formalism by \citet{2005beck}. This approach of applying both formalisms to estimate the magnetic field has also been used in other studies \citep[e.g.][]{2022pandge, 2022dabhade}. The revised formalism by \citet{2005beck} takes into account the value of minimum Lorentz factor ($\gamma_{\mathrm{min}}$). Using a value of $\gamma_{\mathrm{min}}$ = 100, we have calculated the magnetic field using the following modified expression:

\begin{equation}
    B_{\mathrm{eq}}^{'} = 1.1\gamma_{\mathrm{min}}^{\frac{1-2\alpha}{3+\alpha}}B_{\mathrm{eq}}^{\frac{7}{2(3+\alpha)}}
    \label{eq:magfield-eqmod}
\end{equation}

Using the above expressions, the magnetic field strength of J1051+5523 is estimated to be $\sim 1.18\,\mu\mathrm{G}$ according to the classical formalism and $\sim 1.58\,\mu\mathrm{G}$ according to the revised formalism. Similarly, we also estimated the minimum energy densities and magnetic fields of the northern and southern lobes. Using the TGSS radio map, the source depth and surface brightness were measured at a frequency of $\nu_\mathrm{{0}}$ = 150 MHz, for both the northern and southern lobes. The spectral indices for the lobes were estimated from their spectra, as given in Table~\ref{tab:lobe-flux-density}. A summary of the magnetic field estimates for the lobes is provided in Table~\ref{tab:lobe-spectral-age}. Our estimates of the lobe magnetic fields are consistent with typical values of magnetic fields for the lobes of radio galaxies reported in the literature \citep[e.g.][]{2005croston}.

\begin{table}
    \centering
    \caption{Flux density values of the northern and southern lobes of J1051+5523 at different frequencies. Measurements were  obtained using the {\tt CASA} task {\tt imview} from the archival TGSS and NVSS data, along with our uGMRT Band 3 and 4 observations. The regions used for extracting the lobe flux densities are overlaid on the Band 3 image and are provided in Fig.~\ref{fig:lobe-fluxdensity-regions}. The same regions are used for extracting the lobe flux densities from Band 4, TGSS, and NVSS radio images. The flux density of the two point sources near the southern lobe are subtracted as mentioned in Section~\ref{sec:flux-density} and Section~\ref{sec:specindex}.}
	\label{tab:lobe-flux-density}
	\begin{tabular}{lcccc} 
		\hline
		Lobe & $S_\mathrm{{150 MHz}}$ & $S_\mathrm{{402 MHz}}$ & $S_\mathrm{{648 MHz}}$ & $S_\mathrm{{1400 MHz}}$\\
                & (Jy) & (Jy) & (Jy) & (Jy) \\
		\hline
		Northern & 0.65 $\pm$ 0.13 & 0.43 $\pm$ 0.04 & 0.34 $\pm$ 0.03 & 0.18 $\pm$ 0.005 \\
            Lobe & & & & \\
            Southern & 0.64 $\pm$ 0.13 & 0.37 $\pm$ 0.04 & 0.30 $\pm$ 0.03 & 0.13 $\pm$ 0.004 \\
            Lobe & & & & \\
		\hline
	\end{tabular}
\end{table}

With information about the magnetic field and radio spectrum, we can estimate the spectral age of the radio source. The spectral age of a source is defined as the time elapsed since the last acceleration of the synchrotron particles. It can be determined from the steepening of the observed radio spectrum, which result from synchrotron and inverse Compton (IC) loss processes \citep[e.g.][]{1973jaffe, 1999murgia, 2008jamrozy}. The shape of the spectra is fundamentally determined by two factors: the population of electrons experiencing synchrotron losses and the injection of fresh high-energy electrons. High-energy electron losses cause the spectrum to steepen beyond a frequency known as the break frequency ($\nu_{\mathrm{b}}$). In the absence of fresh electron injection or re-acceleration, this steepening becomes more pronounced. Using the known break frequency and the magnetic field, the spectral age of the source can be determined with the following relation provided in \citet{2018turner} as:

\begin{equation}
    \tau = \frac{\nu B^{1/2}}{B^{2}+B_{\mathrm{ic}}^{2}}\left[\nu_{\mathrm{b}} (1+z)\right]^{-1/2}
    \label{eq:spectral-age}
\end{equation}

where $B$ is the equipartition magnetic field, $\nu_{\mathrm{b}}$ is the break frequency, $z$ is the redshift of the source, and $B_{\mathrm{ic}}$ = 0.318$(1+z)^2$ nT is the magnitude of the magnetic field equivalent to the microwave background. The constant of proportionality $\nu$ is given by equation~(\ref{eq:spectral-age-constant}). 

\begin{equation}
    \nu = \left(\frac{243 \pi m_\mathrm{{e}}^{5}c^{2}}{4\mu_\mathrm{{0}}^{2}e^{7}}\right)^{1/2}
    \label{eq:spectral-age-constant}
\end{equation}

where $\mu_\mathrm{{0}}$ is the magnetic permeability of free space.

\begin{figure}
    \includegraphics[width=\columnwidth]{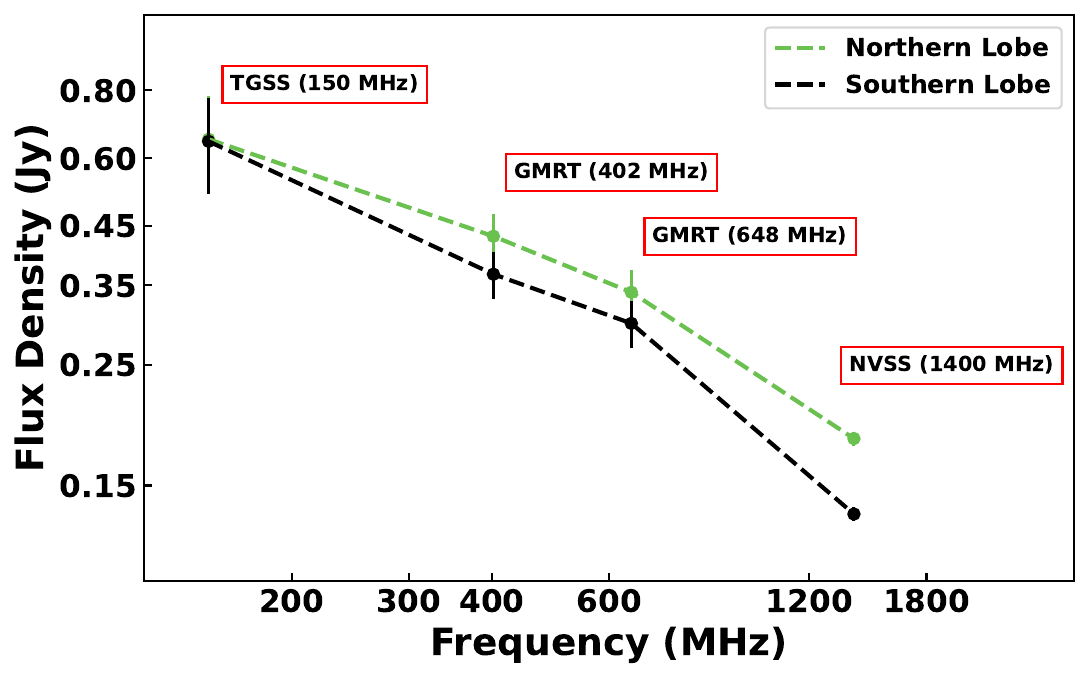}
            \caption{The spectra of the northern and southern lobes of J1051+5523. In both the spectra, a possible spectral break is seen at 648 MHz, which is then considered as the break frequency.}
    \label{fig:lobe-spectral-age}
\end{figure}

In the basic spectral aging model, the magnetic field is assumed to remain constant within the source region. Furthermore, the synchrotron particles are assumed to maintain a uniform power-law energy spectrum. The spectra of the north and south lobes shown in Fig.~\ref{fig:lobe-spectral-age}, exhibit a possible break in the range 650-1400 MHz. The lobe spectra are fitted with both single power-law and broken power-law models and the Bayesian Information Criterion (BIC) was calculated, to estimate the significance of the break in each case. For the southern lobe, the BIC difference between the two models was 1.41, suggesting a possible break with modest statistical significance at 660 MHz. In the northern lobe, the BIC values are very close for both the models, indicating no strong evidence for a break, although its presence cannot be ruled out. The broken power-law fit for the northern lobe yields a break at 590 MHz. Based on these results, we assumed our Band 4 frequency (648 MHz) as the break frequency, $\nu_\mathrm{{b}}$. Using equation~\ref{eq:spectral-age}, we then estimated the spectral ages of the individual lobes. A summary of the spectral age estimates is provided in Table~\ref{tab:lobe-spectral-age}. From our estimates, it is seen that the northern and southern lobes have similar spectral ages of $\sim$150 Myr.

The spectral age estimated for J1051+5523 is comparable to that of WAT radio galaxies residing in clusters \citep[e.g.][]{2008giacintucci}. The spectral age of $\sim$150 Myr for the radio lobes, combined with the projected linear size of $\sim$488 kpc, suggests that the lobes have expanded at a sub-relativistic speed. As a WAT radio source hosted by a BCG in a cluster, such low expansion speeds are expected \citep{2004hardcastle}. The relatively long spectral age of the lobes implies that they have evolved slowly, probably due to confinement by the dense ICM at the cluster core, thereby reducing adiabatic losses and maintaining the synchrotron electron population over extended timescales \citep{2011murgia}.

The presence of an active radio core in the radio maps suggests ongoing AGN activity. However, the radio maps provide no insight into episodic AGN activity. We find no morphological evidence of older outer lobes or reoriented jets that typically characterise episodic sources \citep[e.g.][]{2009saikia}. Recent studies by \citet{2025raj} have proposed a method in which a concave curvature in the integrated spectrum could be considered as a probe of episodic activity, even if morphological evidence is absent. However, the spectrum of J1051+5523 did not exhibit concave curvature, which limits the possibility of checking for episodic activity. In this context, a plausible evolutionary scenario is one of long-lived, continuous activity, where the AGN has steadily supplied energy to the lobes for $\sim$150 Myr. This evolutionary picture aligns well with other WATs found in cluster centres, which tend to exhibit slow-growing lobes shaped by ram pressure and long-term AGN feedback \citep{2018hardcastle}. A single AGN episode lasting $\sim$50-200 Myr within a cluster environment has been explored in previous studies \citep[e.g.][]{2008jetha}. Furthermore, the studies by \citet{2012antognini} estimate that a typical timescale of $\sim$190 Myr for continuous activity is plausible in cluster-central radio galaxies hosted by BCGs, supporting the consistency of our results with those observed in similar environments.

\begin{table*}[h]
	\centering
	\caption{Spectral age estimates of J1051+5523. The spectral index, $\alpha$ in column 3 is estimated by fitting straight lines on the spectra given in Fig.~\ref{fig:lobe-spectral-age}. The magnetic field values in column 6 corresponds to the revised formalism and the corresponding spectral ages are given in column 8.}
	\label{tab:lobe-spectral-age}
	\begin{tabular}{lccccccc}
	\hline
	Region & depth & $\mathrm{\alpha}$ & $\mathrm{\nu_{b}}$ & $\mathrm{B_{eq}}$ & $\mathrm{B_{eq}^{'}}$ & $\mathrm{\tau}$ & $\mathrm{\tau^{'}}$\\
         & (kpc) & & (MHz) & $\mathrm{\mu}$G & $\mathrm{\mu}$G & Myr & Myr \\
        (1) & (2) & (3) & (4) & (5) & (6) & (7) & (8)\\
	\hline
        Northern Lobe & 103.19 & -0.57 & 648 & 1.34 & 1.61 & 144.75 & 150.65 \\
        Southern Lobe & 97.1 & -0.71 & 648 & 1.35 & 1.89 & 144.83 & 153.86 \\
	\hline
	\end{tabular}
\end{table*}

\subsection{Jet and Lobe Spectral Index}
\begin{figure}
    \includegraphics[width=\columnwidth]{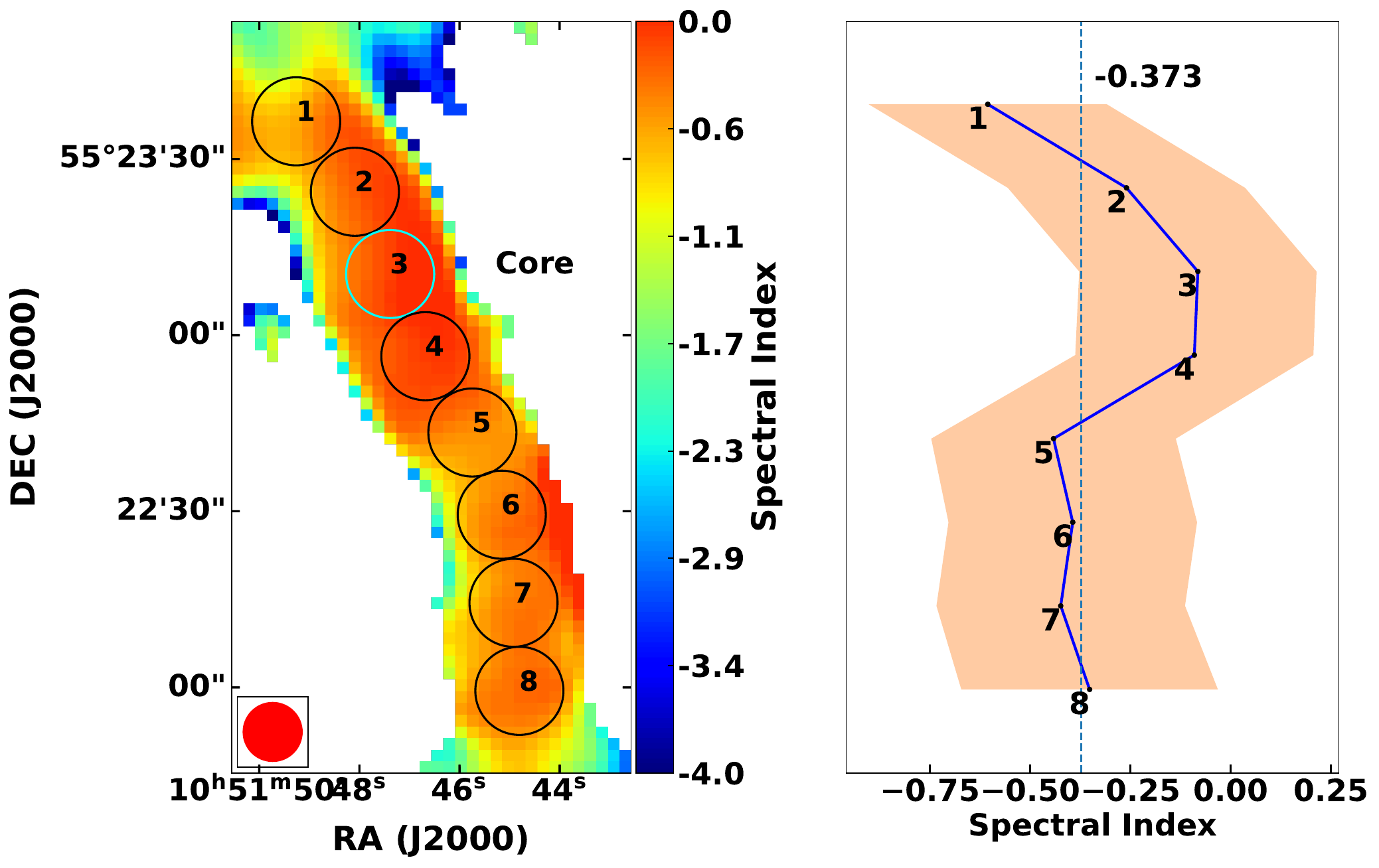}
    \caption{Left panel: Spectral index map of the radio jet. The black circular regions are drawn with diameter 1.5 times the beam size. The radio beam is provided at the bottom left corner of the figure. The radio core is shown with a cyan circle. Right panel: Spectral index variation along the length of the jet. The blue dashed line shows the median spectral index value. The shaded region shows the mean error in spectral index obtained for each region from the spectral index error map, together with a calibration error of 10\% added in quadrature.}
    \label{fig:jet-spectral-index}
\end{figure}

To check the variation of the spectral index along the jets and lobes, we selected different regions and estimated the mean spectral index in each region and plotted their variation along the regions. The spatially resolved spectral index map shows localised features. However, there exists the possibility of local noise variations within the spectral index map. Averaging the spectral index within a specific region helps to suppress this local noise in the spectral index and extract meaningful trends. To plot the spectral index variation along the radio jet, we drew circular regions with a diameter of 1.5 times the beam size of the spectral index map. The mean spectral index in each circular region was then estimated and plotted along the length of the radio jet. The plot is presented in Fig.~\ref{fig:jet-spectral-index}. The core exhibits the flattest spectral index of -0.082. However, the jet lengths are not symmetric, with the southern jet being relatively longer than the northern jet. The asymmetry in jet length could be caused by environmental effects or intrinsic differences in the jet propagation, or it could be due to possible orientation effects. Moving from the core along the northern jet, the spectral index steepens. Similarly, along the southern jet, the spectral index steepens up to a point, after which it plateaued at around -0.4. The plateau could suggest an ongoing or recent re-acceleration of particles or some interaction with the external medium, slowing the synchrotron energy losses. The median spectral index along the jet is observed to be -0.373.

\begin{figure*}
    \includegraphics[width=7in]{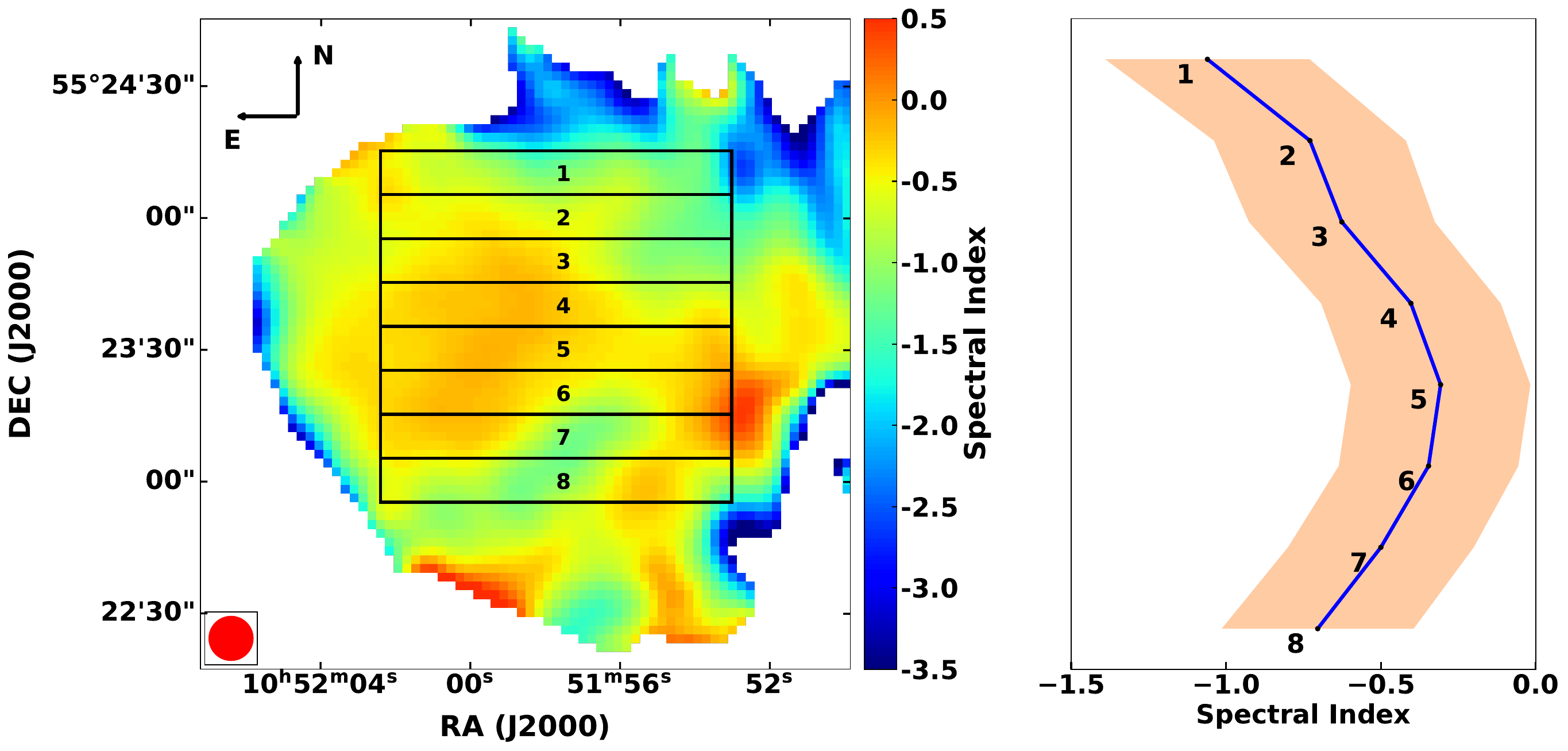}
    \caption{Left panel: Spectral index map of the northern lobe. The black rectangular regions are drawn with width of the beam size and length 8 times the beam size. The radio beam is provided at the bottom left corner of the figure. Top left corner shows the directions. Right panel: Spectral index variation along different regions of the northern lobe. The shaded region shows the mean error in spectral index obtained for each region from the spectral index error map, together with a calibration error of 10\% added in quadrature.}
    \label{fig:northern-lobe-spectral-index}
\end{figure*}

The northern jet starts from the core and, after traversing a distance of $\sim$48 kpc, bends towards the eastern direction. The northern lobe has a region with a relatively flat spectral index. The high-resolution radio maps show high-surface-brightness emission from this region. It could be the freshly injected synchrotron particles by the northern jet. Conversely, being a cluster center source, the enhanced surface brightness in the northern lobe may also result from the interaction with the ICM, leading to localised particle re-acceleration. The dense environment could also be suppressing synchrotron and inverse-Compton losses, thus slowing the energy loss of the plasma, which could also result in a flat spectrum region. In order to get an average profile of the spectral index from the lobe, we drew rectangular regions with their lengths along the east-west direction. The width of the regions was chosen to be the beam size, and the length of the regions was chosen to be eight times the beam size to ensure optimum coverage of the lobe area. The mean spectral index in each rectangular region was then estimated, and its variation across different regions was plotted. The plot is presented in Fig.~\ref{fig:northern-lobe-spectral-index}. The average spectral index profile of the northern lobe shows a few regions (regions 4, 5, and 6), with a relatively flat spectral index ($\sim$ –0.3). This is not expected for optically thin plasma in the lobe. The uncertainties in our brightness estimates may contribute to this flattening. With a calibration uncertainty of 10\%, we estimate the average spectral index of region 5 to be -0.30 $\pm$ 0.29. This results in a lower limit of -0.59, which is consistent with freshly accelerated particles. In addition, regions 4–6 exist in the region where the northern jet appears to terminate and inject fresh relativistic particles into the lobe. This region is likely to be the site of jet-lobe interaction, where localised re-acceleration due to jet impact could lead to a relatively flatter spectrum. However, no hotspots indicative of a termination shock are observed. The lobe is seen to expand laterally from this zone on either side. The gradual steepening of the spectral index towards the lateral sides implies synchrotron aging as electrons propagate and lose energy away from the injection site (regions 4–6). The smooth gradient of the spectral index and its symmetrical profile along the regions support continuous particle transport and aging within the lobe, without evidence for significant re-acceleration in the outer regions. But, in order to have an accurate picture about the flat nature of the lobe spectrum, deeper observations and independent measurements are required.

\begin{figure*}
    \includegraphics[width=7in]{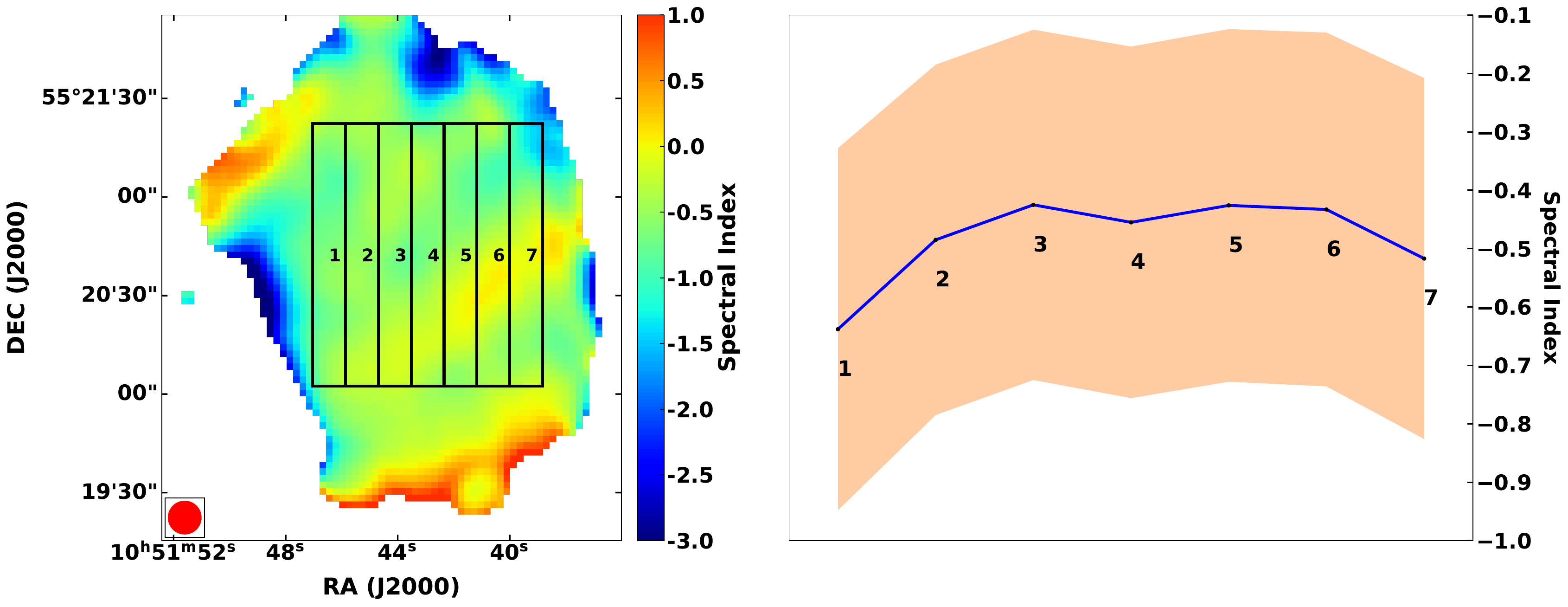}
    \caption{Left panel: Spectral index map of the southern lobe. The black rectangular regions are drawn with width of the beam size and height 8 times the beam size. The radio beam is provided at the bottom left corner of the figure. Right panel: Spectral index variation along different regions of the southern lobe. The shaded region shows the mean error in spectral index obtained for each region from the spectral index error map, together with a calibration error of 10\% added in quadrature.}
    \label{fig:southern-lobe-spectral-index}
\end{figure*}

The southern jet starts from the core and, after traversing a distance of $\sim$197 kpc, connects to the southern lobe. The southern lobe shows a few structures with sharp variations in the spectral index between them. The Band 3 and Band 4 radio maps show relatively low surface brightness, cavity-like regions within the southern lobe, interspaced by horizontal structures. The southern jet also seems to traverse all the way down to the bottom of the lobe. The relatively flat horizontal structures seen in the spectral index map are on the top and bottom sides of the cavity-like regions, which themselves are seen as relatively steep. From the high-resolution radio maps, we infer that the observed structures in the southern lobe are formed by the backflow of radio plasma from its termination points. However, no terminal hotspot is observed. \citet{2015kolokythas} discussed a scenario in which backflow within the lobes transports young, energetic electrons from the jet into regions of older plasma, potentially creating spectral index structures. We also suspect the possibility of beam smearing or projection effects, which could also result in sharp transitions in the spectral index among the structures. For the southern lobe also, we tried to plot an average spectral index profile. In the southern lobe, the regions were chosen to be vertical, with a height eight times the beam size and a width equal to the beam size. The mean spectral index in each region is then estimated. The variation of this spectral index along the regions is plotted in Fig.~\ref{fig:southern-lobe-spectral-index}. The spectral index remains within a limited range, not showing considerable variation compared to the northern lobe. Across regions 2–6, the spectral index is not too steep ($\alpha \sim$ –0.4), which may suggest recent or ongoing particle injection by the southern jet. Further, it could also suggest a broader jet–lobe interaction region, when compared to the northern lobe. In the southern lobe also, the spectral index profile suggests a lateral propagation of the electrons from the jet–lobe interaction region. However, there is no considerable variation in the spectral profile along the regions, which implies a more homogeneous radio plasma in the southern lobe. This is an indication that the lobe could be young, which is, however, not observed from our spectral age estimates. The next possible explanation could be that the synchrotron energy losses are balanced by mild re-acceleration or mixing, thereby maintaining spectral uniformity.

\section{Discussion}
\label{sec:discussions}
J1051+5523 is a prototypical WAT radio galaxy associated with the galaxy cluster WHL J105147.4+55230. The host galaxy is identified as a BCG within the cluster and further classified as a cD galaxy. The radio power of J1051+5523 at 150 MHz, as determined from the TGSS radio map, is $2.91 \times 10^{25}\,\mathrm{W\,Hz^{-1}}$. According to \citet{2019mingo}, the lower limit of radio power at 150 MHz for FR II sources is $\sim 10^{26}\,\mathrm{W\,Hz^{-1}}$. Since the estimated power is below this lower limit, J1051+5523 is classified as a source in the transitional regime between FR I and FR II, which is where the WATs are found according to \citet{2019missaglia}. However, while the considerations on radio power may be valid, the morphology of J1051+5523 is not really compatible with a relativistic jet, as expected in FRIIs.

J1051+5523 exhibits a distinctive bend in its northern jet, nearly at a right angle, and its overall morphology resembles an axe. Sources with similar abrupt bends have been observed previously. The radio galaxies 1919+479, as described by \citet{1981burns}, 0836+290 as described by \citet{1986burns}, and 2236-176, as noted by \citet{1993odonoghue} and \citet{2004hardcastle}, exhibit a morphology similar to that of J1051+5523. In such sources, compact regions known as warmspots are typically observed in areas where jet-plume transitions occur. However, no such warmspots are observed in J1051+5523. Further, the above-mentioned radio sources are all showing extensive plumes or tails after the abrupt bend, which is not observed in J1051+5523. Here, the northern jet bends abruptly towards the east and continues without disruption, until it expands into a small extended structure without a compact warmspot. Such an abrupt bent indicates that the jet stability is suddenly disrupted \citep[e.g.][]{1981burns}. \citet{1986burns} proposed that the bends and asymmetries observed in similar source structures could be due to the radio jets being impacted by an external impulsive force at irregular distances from the cluster core. This impulsive force could potentially result from interactions between the radio jets and clouds in the ICM. Such asymmetries can also be caused by recollimation shocks wherein the initially expanding jet of plasma is squeezed inward by external pressure, causing it to recollimate or change the jet direction.

In the case of the southern jet, a slight kink is observed, after which the surface brightness decreases, but the jet continues without further disruption until it flares into a small, lobe-like structure, which lacks compact warmspots. Additionally, high-resolution uGMRT radio maps reveal a jet-like structure with multiple kinks within the expanded region of the southern lobe. The lack of compact warmspots in the jet-plume transition region is evidenced by a gradual transition from the jet into the plume base, with the jet remaining well-collimated for a period after entering the plume \citep{2004hardcastle}. In J1051+5523, we are not observing relatively steep lobes with the northern lobe having a spectral index, $\alpha_\mathrm{{1400}}^\mathrm{{150}}$ = -0.57 and the southern lobe having a spectral index, $\alpha_\mathrm{{1400}}^\mathrm{{150}}$ = -0.71. However, we observe certain structures in the two lobes from the spectral index map. The northern lobe contains a region with a relatively flat spectral index, which could be due to ongoing particle injection from the northern jet. This could be further enhanced by interactions with the dense ICM around the northern lobe, which may promote localised re-acceleration and suppress energy losses, maintaining a flatter spectrum. Meanwhile, the horizontal structures seen in the spectral index map of the southern lobe suggest a complex plasma flow, probably shaped by backflow from the jet termination, though the absence of a clear hotspot also suggests the possibility of projection effects or beam smearing.

As mentioned previously, the radio galaxy 1919+479 exhibits a morphology similar to that of J1051+5523. \citet{1994pinkney} estimated a very small peculiar velocity for 1919+479, with a value of < 50 $\mathrm{km\,s^{-1}}$. They also classified the associated cluster as poor and noted its less dense environment. Through substructure analysis, they proposed a merger scenario related to the cluster, which likely led to the formation of 1919+479. For J1051+5523, we estimated the relative velocity of the host galaxy. From \citet{2023kim}, the velocity of the galaxy is found to be 22164 $\pm$ 5 $\mathrm{km\,s^{-1}}$. The estimate is based on spectroscopic SDSS data. For the cluster, \citet{2021kirkpatrick} estimated a velocity of 21886 $\pm$ 2643 $\mathrm{km\,s^{-1}}$, which is based on photometric redshift values. The uncertainty associated with the cluster velocity is large. Based on the velocity estimates, the relative velocity of the galaxy is obtained to be 278 $\pm$ 2643 $\mathrm{km\,s^{-1}}$. Given this significant uncertainty, we can neither ascertain nor rule out the possibility of a substantial relative motion for the galaxy, which could contribute to the observed WAT morphology. Also, these are radial velocity estimates. If the galaxy has relative motion in the plane of the sky, there will be transverse velocity components, which we cannot measure directly. The bending of the northern jet may suggest such transverse motion as a plausible cause. However, to obtain a robust estimate of the relative velocity and to check if it is sufficient to generate the ram pressure required for bending the jets, accurate spectroscopic data are required for the cluster. 

As the next possible cause, we discuss buoyancy as a potential explanation for the observed bend. Asymmetric buoyant motion of radio lobes has been observed in various WAT systems \citep[e.g.][]{1984robertson}. Another bent source with similar morphology was studied by \citet{1982burnsjo}. Using X-ray data, the authors proposed that the morphology of the radio galaxy 1159+583 could be due to a combination of buoyancy and a dynamic pressure gradient caused by the host galaxy moving at a relative velocity of $\sim$200 $\mathrm{km\,s^{-1}}$, with respect to the host cluster. They also suggested that the ICM gas not only influences the shape of the radio emission but also exerts pressure that counteracts adiabatic expansion losses in the radio structure. In the case of J1051+5523, the lobe's spectral age of $\sim$150 Myr, along with its relatively less steep spectral index, could be attributed to such suppressed energy losses under such environments. \citet{1982burnsjo} also observed that the host cluster has a smooth, regular X-ray profile, which indicates an unrelaxed cluster in its early evolutionary stage. In an alternative study, in a poor cluster, with evidence of a merger, \citet{1996sakelliou} observed a WAT source with a relative velocity of $\sim$300 $\mathrm{km\,s^{-1}}$. For a galaxy with relative velocity in this range, the authors explained the morphology as a result of both the ram pressure and buoyancy forces. In the case of J1051+5523, assuming a lower limit for the relative velocity of 278 $\mathrm{km\,s^{-1}}$ within the uncertainty limits, is not sufficient to produce the required ram pressure to explain the observed bent morphology. Moreover, since the host galaxy is a BCG located within the gravitational potential of the cluster, the likelihood of it having a high relative velocity is low. However, because of the large uncertainty, this cannot be confirmed. There also exist various other models for jet bending, including the cloud-collision model \citep{1986burns}, the adiabatic model, and the kinetic model \citep{1993odonoghue}, among others. However, evaluating the applicability of these models is beyond the scope of this work.

We estimated the mass of the cluster to be $2.03 \times 10^{14}\,M_\odot$ as follows \citet{2012wen}. The relation used for estimating the mass is given in Equation~\ref{eq:whl2012}.
\begin{equation}
 \log M_{200} = (-1.49 \pm 0.05) + (1.17 \pm 0.03)\,\log R_{\mathrm{L*}}
    \label{eq:whl2012}
\end{equation}

\noindent where $R_{\mathrm{L*}}$ is the richness of the galaxy cluster. From \citet{2012wen}, $R_{\mathrm{L*}}$ is found to be 34.45. The estimated mass shows that the cluster hosting J1051+5523 is a low-mass cluster. The region is observed by the Chandra X-ray telescope (Obs ID: 11590, exposure time: 21.85 ks). However, only the central AGN is visible in the X-ray image, with no evidence of X-ray emission from the surrounding cluster environment. The low cluster mass together with the absence of ambient X-ray emission points to the absence of any major merger events, which could result in WAT morphologies. This is further suggested in the studies by \citet{2017tempel}, where the authors were unable to find the cluster to be a merging system. However, they used a modified Friend-of-Friend (FoF) algorithm, which is automated and thus it may not be a straightforward method to identify merger systems. Hence we cannot rule out the possibility of a minor merger which could potentially result in the morphology of J1051+5523. Recollecting all the above mentioned concepts, we suggest that the observed morphology of J1051+5523 may result from a combination of ram pressure and/or buoyant forces within the cluster environment. However, since we cannot firmly confirm the role of the cluster environment in shaping the structure of J1051+5523, the alternative possibility that the observed morphology arises from the projection of a weakly bent C-shaped radio galaxy inclined to the plane of the sky cannot be entirely excluded.

\section{Conclusions}
\label{sec:conclusion}
We have presented the multi-frequency study of an axe-shaped bent tail radio galaxy in a cluster. J1051+5523 is an archetype wide-angle tail galaxy in the galaxy cluster WHL J105147.4+552309. Deep, high-resolution uGMRT radio maps show the intricate radio morphology of the source with a nearly right-angled bent of the northern jet and multiple kinks in the southern jet. The estimated radio power of $2.91 \times 10^{25}\,\mathrm{W\,Hz^{-1}}$ at 150 MHz indicates that J1051+5523 lies in the transition region between FRIs and FRIIs, where wide-angle tailed sources are generally seen. The spectral index map shows a flat core region and relatively flat lobes, which may indicate an ongoing particle acceleration or a relatively young population of relativistic electrons in the lobes. Using multi-frequency radio data, we estimated the equipartition magnetic fields and spectral ages of the lobes as well. There is weak evidence for a break in the spectrum, which would correspond to an age of 150 Myr for the lobes. The estimates suggest a long-lived radio source with continuous activity, where the AGN has steadily supplied energy to the lobes for $\sim$150 Myr. We estimated a relative velocity of 278 $\pm$ 2643 $\mathrm{km\,s^{-1}}$ for the host galaxy with respect to the cluster. Since there exists a large uncertainty associated with the relative velocity estimates, it is difficult to determine whether the velocity is sufficient or not to generate the ram pressure required to bend the jets. Also, the bending of the northern jet may suggest that a transverse motion of the galaxy could be a plausible cause, which, however, cannot be measured directly. We also explored the possibility of buoyancy forces which could aid in bending the jet. The host cluster is also estimated to have a low mass of $\sim$2$\times 10^{14}$~$M_\odot$. The low cluster mass, together with the absence of diffuse X-ray emission, may suggest the absence of any major mergers associated with the cluster, which could result in WAT morphology. In summary, we suggest that the observed WAT morphology of J1051+5523 is possibly formed from a combination of ram pressure and/or buoyant forces within the cluster environment. Further observations in X-ray as well as polarisation studies are required to confirm the exact mechanism of bending, the cluster gas distribution, and cluster dynamics.


\section*{Acknowledgements}
We thank the anonymous reviewer for their suggestions and comments. C. H. Ishwara-Chandra, RK, and RS acknowledges the support of the Department of Atomic Energy, Government of India, under project no. 12-R\&D-TFR-5.02-0700. VJ acknowledges the support provided by the Department of Science and Technology (DST) under the ‘Fund for Improvement of S \& T Infrastructure (FIST)’ program (SR/FST/PS-I/2022/208). VJ, and JJ also thanks the Inter-University Centre for Astronomy and Astrophysics (IUCAA), Pune, India, for the Visiting Associateship. We thank the staff of the GMRT that made these observations possible. The GMRT is run by the National Centre for Radio Astrophysics (NCRA) of the Tata Institute of Fundamental Research (TIFR). This research made use of Astropy \footnote{\url{https://www.astropy.org/}} a community-developed core Python package for Astronomy \citep{2013astropy}, and APLpy, an open-source plotting package for Python \citep{2012robitaille}.

\section*{Data Availability}

The reduced radio data and images underlying this article will be shared on reasonable request to the corresponding author. The raw uGMRT data used in this work are already available in the GMRT archive (\url{https://naps.ncra.tifr.res.in/goa/data/search}). Moreover, the NVSS data used in this work are available in the NVSS archive (\url{https://www.cv.nrao.edu/nvss/}), and TGSS data are available in the TGSS archive (\url{https://vo.astron.nl/tgssadr/q_fits/cutout/form}).



\bibliographystyle{mnras}
\bibliography{sudheesh} 

\begin{thebibliography}{}
\makeatletter
\relax
\def\mn@urlcharsother{\let\do\@makeother \do\$\do\&\do\#\do\^\do\_\do\%\do\~}
\def\mn@doi{\begingroup\mn@urlcharsother \@ifnextchar [ {\mn@doi@} {\mn@doi@[]}}
\def\mn@doi@[#1]#2{\def\@tempa{#1}\ifx\@tempa\@empty \href {http://dx.doi.org/#2} {doi:#2}\else \href {http://dx.doi.org/#2} {#1}\fi \endgroup}
\def\mn@eprint#1#2{\mn@eprint@#1:#2::\@nil}
\def\mn@eprint@arXiv#1{\href {http://arxiv.org/abs/#1} {{\tt arXiv:#1}}}
\def\mn@eprint@dblp#1{\href {http://dblp.uni-trier.de/rec/bibtex/#1.xml} {dblp:#1}}
\def\mn@eprint@#1:#2:#3:#4\@nil{\def\@tempa {#1}\def\@tempb {#2}\def\@tempc {#3}\ifx \@tempc \@empty \let \@tempc \@tempb \let \@tempb \@tempa \fi \ifx \@tempb \@empty \def\@tempb {arXiv}\fi \@ifundefined {mn@eprint@\@tempb}{\@tempb:\@tempc}{\expandafter \expandafter \csname mn@eprint@\@tempb\endcsname \expandafter{\@tempc}}}

\bibitem[\protect\citeauthoryear{{Ahumada} et~al.,}{{Ahumada} et~al.}{2020}]{2020ahumada}
{Ahumada} R.,  et~al., 2020, \mn@doi [\apjs] {10.3847/1538-4365/ab929e}, \href {https://ui.adsabs.harvard.edu/abs/2020ApJS..249....3A} {249, 3}

\bibitem[\protect\citeauthoryear{{Antognini}, {Bird}  \& {Martini}}{{Antognini} et~al.}{2012}]{2012antognini}
{Antognini} J.,  {Bird} J.,   {Martini} P.,  2012, \mn@doi [\apj] {10.1088/0004-637X/756/2/116}, \href {https://ui.adsabs.harvard.edu/abs/2012ApJ...756..116A} {756, 116}

\bibitem[\protect\citeauthoryear{{Astropy Collaboration} et~al.,}{{Astropy Collaboration} et~al.}{2013}]{2013astropy}
{Astropy Collaboration} et~al., 2013, \mn@doi [\aap] {10.1051/0004-6361/201322068}, \href {https://ui.adsabs.harvard.edu/abs/2013A&A...558A..33A} {558, A33}

\bibitem[\protect\citeauthoryear{{Beck} \& {Krause}}{{Beck} \& {Krause}}{2005}]{2005beck}
{Beck} R.,  {Krause} M.,  2005, \mn@doi [Astronomische Nachrichten] {10.1002/asna.200510366}, \href {https://ui.adsabs.harvard.edu/abs/2005AN....326..414B} {326, 414}

\bibitem[\protect\citeauthoryear{{Becker}, {White}  \& {Edwards}}{{Becker} et~al.}{1991}]{1991becker}
{Becker} R.~H.,  {White} R.~L.,   {Edwards} A.~L.,  1991, \mn@doi [\apjs] {10.1086/191529}, \href {https://ui.adsabs.harvard.edu/abs/1991ApJS...75....1B} {75, 1}

\bibitem[\protect\citeauthoryear{{Blandford} \& {Rees}}{{Blandford} \& {Rees}}{1974}]{1974blandford}
{Blandford} R.~D.,  {Rees} M.~J.,  1974, \mn@doi [\mnras] {10.1093/mnras/169.3.395}, \href {https://ui.adsabs.harvard.edu/abs/1974MNRAS.169..395B} {169, 395}

\bibitem[\protect\citeauthoryear{{Bodo}, {Ferrari}, {Massaglia}  \& {Tsinganos}}{{Bodo} et~al.}{1985}]{1985bodo}
{Bodo} G.,  {Ferrari} A.,  {Massaglia} S.,   {Tsinganos} K.,  1985, \aap, \href {https://ui.adsabs.harvard.edu/abs/1985A&A...149..246B} {149, 246}

\bibitem[\protect\citeauthoryear{{Burbidge}}{{Burbidge}}{1956}]{1956burbidge}
{Burbidge} G.~R.,  1956, \mn@doi [\apj] {10.1086/146237}, \href {https://ui.adsabs.harvard.edu/abs/1956ApJ...124..416B} {124, 416}

\bibitem[\protect\citeauthoryear{{Burns}}{{Burns}}{1981}]{1981burns}
{Burns} J.~O.,  1981, \mn@doi [\mnras] {10.1093/mnras/195.3.523}, \href {https://ui.adsabs.harvard.edu/abs/1981MNRAS.195..523B} {195, 523}

\bibitem[\protect\citeauthoryear{{Burns}}{{Burns}}{1986}]{1986burns}
{Burns} J.~O.,  1986, \mn@doi [Canadian Journal of Physics] {10.1139/p86-065}, \href {https://ui.adsabs.harvard.edu/abs/1986CaJPh..64..373B} {64, 373}

\bibitem[\protect\citeauthoryear{{Burns} \& {Balonek}}{{Burns} \& {Balonek}}{1982}]{1982burnsjo}
{Burns} J.~O.,  {Balonek} T.~J.,  1982, \mn@doi [\apj] {10.1086/160525}, \href {https://ui.adsabs.harvard.edu/abs/1982ApJ...263..546B} {263, 546}

\bibitem[\protect\citeauthoryear{{Burns} \& {Owen}}{{Burns} \& {Owen}}{1980}]{1980burns}
{Burns} J.~O.,  {Owen} F.~N.,  1980, \mn@doi [\aj] {10.1086/112663}, \href {https://ui.adsabs.harvard.edu/abs/1980AJ.....85..204B} {85, 204}

\bibitem[\protect\citeauthoryear{{Burns}, {Eilek}  \& {Owen}}{{Burns} et~al.}{1982}]{1982burns}
{Burns} J.~O.,  {Eilek} J.~A.,   {Owen} F.~N.,  1982, in {Heeschen} D.~S.,  {Wade} C.~M.,  eds, ~ Vol. 97, Extragalactic Radio Sources. p.~45

\bibitem[\protect\citeauthoryear{{Capetti}, {Massaro}  \& {Baldi}}{{Capetti} et~al.}{2017}]{2017capetti}
{Capetti} A.,  {Massaro} F.,   {Baldi} R.~D.,  2017, \mn@doi [\aap] {10.1051/0004-6361/201629287}, \href {https://ui.adsabs.harvard.edu/abs/2017A&A...598A..49C} {598, A49}

\bibitem[\protect\citeauthoryear{{Chandra} \& {Kanekar}}{{Chandra} \& {Kanekar}}{2017}]{2017chandra}
{Chandra} P.,  {Kanekar} N.,  2017, \mn@doi [\apj] {10.3847/1538-4357/aa85a2}, \href {https://ui.adsabs.harvard.edu/abs/2017ApJ...846..111C} {846, 111}

\bibitem[\protect\citeauthoryear{{Condon}, {Cotton}, {Greisen}, {Yin}, {Perley}, {Taylor}  \& {Broderick}}{{Condon} et~al.}{1998}]{1998condon}
{Condon} J.~J.,  {Cotton} W.~D.,  {Greisen} E.~W.,  {Yin} Q.~F.,  {Perley} R.~A.,  {Taylor} G.~B.,   {Broderick} J.~J.,  1998, \mn@doi [\aj] {10.1086/300337}, \href {https://ui.adsabs.harvard.edu/abs/1998AJ....115.1693C} {115, 1693}

\bibitem[\protect\citeauthoryear{{Croston}, {Hardcastle}, {Harris}, {Belsole}, {Birkinshaw}  \& {Worrall}}{{Croston} et~al.}{2005}]{2005croston}
{Croston} J.~H.,  {Hardcastle} M.~J.,  {Harris} D.~E.,  {Belsole} E.,  {Birkinshaw} M.,   {Worrall} D.~M.,  2005, \mn@doi [\apj] {10.1086/430170}, \href {https://ui.adsabs.harvard.edu/abs/2005ApJ...626..733C} {626, 733}

\bibitem[\protect\citeauthoryear{{Dabhade} et~al.,}{{Dabhade} et~al.}{2020}]{2020dabhade}
{Dabhade} P.,  et~al., 2020, \mn@doi [\aap] {10.1051/0004-6361/202038344}, \href {https://ui.adsabs.harvard.edu/abs/2020A&A...642A.153D} {642, A153}

\bibitem[\protect\citeauthoryear{{Dabhade} et~al.,}{{Dabhade} et~al.}{2022}]{2022dabhade}
{Dabhade} P.,  et~al., 2022, \mn@doi [\aap] {10.1051/0004-6361/202243182}, \href {https://ui.adsabs.harvard.edu/abs/2022A&A...668A..64D} {668, A64}

\bibitem[\protect\citeauthoryear{{Donoso}, {Best}  \& {Kauffmann}}{{Donoso} et~al.}{2009}]{2009donoso}
{Donoso} E.,  {Best} P.~N.,   {Kauffmann} G.,  2009, \mn@doi [\mnras] {10.1111/j.1365-2966.2008.14068.x}, \href {https://ui.adsabs.harvard.edu/abs/2009MNRAS.392..617D} {392, 617}

\bibitem[\protect\citeauthoryear{{Edwards}, {Fadda}  \& {Frayer}}{{Edwards} et~al.}{2010}]{2010edwards}
{Edwards} L. O.~V.,  {Fadda} D.,   {Frayer} D.~T.,  2010, \mn@doi [\apjl] {10.1088/2041-8205/724/2/L143}, \href {https://ui.adsabs.harvard.edu/abs/2010ApJ...724L.143E} {724, L143}

\bibitem[\protect\citeauthoryear{{Falle}}{{Falle}}{1991}]{1991falle}
{Falle} S.~A.~E.~G.,  1991, \mn@doi [\mnras] {10.1093/mnras/250.3.581}, \href {https://ui.adsabs.harvard.edu/abs/1991MNRAS.250..581F} {250, 581}

\bibitem[\protect\citeauthoryear{{Fanaroff} \& {Riley}}{{Fanaroff} \& {Riley}}{1974}]{1974fanaroff}
{Fanaroff} B.~L.,  {Riley} J.~M.,  1974, \mn@doi [\mnras] {10.1093/mnras/167.1.31P}, \href {https://ui.adsabs.harvard.edu/abs/1974MNRAS.167P..31F} {167, 31P}

\bibitem[\protect\citeauthoryear{{Freeland}, {Cardoso}  \& {Wilcots}}{{Freeland} et~al.}{2008}]{2008freeland}
{Freeland} E.,  {Cardoso} R.~F.,   {Wilcots} E.,  2008, \mn@doi [\apj] {10.1086/591443}, \href {https://ui.adsabs.harvard.edu/abs/2008ApJ...685..858F} {685, 858}

\bibitem[\protect\citeauthoryear{{Garon} et~al.,}{{Garon} et~al.}{2019}]{2019garon}
{Garon} A.~F.,  et~al., 2019, \mn@doi [\aj] {10.3847/1538-3881/aaff62}, \href {https://ui.adsabs.harvard.edu/abs/2019AJ....157..126G} {157, 126}

\bibitem[\protect\citeauthoryear{{Giacintucci} et~al.,}{{Giacintucci} et~al.}{2008}]{2008giacintucci}
{Giacintucci} S.,  et~al., 2008, \mn@doi [\apj] {10.1086/589280}, \href {https://ui.adsabs.harvard.edu/abs/2008ApJ...682..186G} {682, 186}

\bibitem[\protect\citeauthoryear{{Govoni} \& {Feretti}}{{Govoni} \& {Feretti}}{2004}]{2004govoni}
{Govoni} F.,  {Feretti} L.,  2004, \mn@doi [International Journal of Modern Physics D] {10.1142/S0218271804005080}, \href {https://ui.adsabs.harvard.edu/abs/2004IJMPD..13.1549G} {13, 1549}

\bibitem[\protect\citeauthoryear{{Gupta} et~al.,}{{Gupta} et~al.}{2017}]{2017gupta}
{Gupta} Y.,  et~al., 2017, \mn@doi [Current Science] {10.18520/cs/v113/i04/707-714}, \href {https://ui.adsabs.harvard.edu/abs/2017CSci..113..707G} {113, 707}

\bibitem[\protect\citeauthoryear{{Hardcastle}}{{Hardcastle}}{2018}]{2018hardcastle}
{Hardcastle} M.~J.,  2018, \mn@doi [\mnras] {10.1093/mnras/stx3358}, \href {https://ui.adsabs.harvard.edu/abs/2018MNRAS.475.2768H} {475, 2768}

\bibitem[\protect\citeauthoryear{{Hardcastle} \& {Croston}}{{Hardcastle} \& {Croston}}{2020}]{2020hardcastle}
{Hardcastle} M.~J.,  {Croston} J.~H.,  2020, \mn@doi [\nar] {10.1016/j.newar.2020.101539}, \href {https://ui.adsabs.harvard.edu/abs/2020NewAR..8801539H} {88, 101539}

\bibitem[\protect\citeauthoryear{{Hardcastle} \& {Sakelliou}}{{Hardcastle} \& {Sakelliou}}{2004}]{2004hardcastle}
{Hardcastle} M.~J.,  {Sakelliou} I.,  2004, \mn@doi [\mnras] {10.1111/j.1365-2966.2004.07522.x}, \href {https://ui.adsabs.harvard.edu/abs/2004MNRAS.349..560H} {349, 560}

\bibitem[\protect\citeauthoryear{{Intema}, {Jagannathan}, {Mooley}  \& {Frail}}{{Intema} et~al.}{2017}]{2017intema}
{Intema} H.~T.,  {Jagannathan} P.,  {Mooley} K.~P.,   {Frail} D.~A.,  2017, \mn@doi [\aap] {10.1051/0004-6361/201628536}, \href {https://ui.adsabs.harvard.edu/abs/2017A&A...598A..78I} {598, A78}

\bibitem[\protect\citeauthoryear{{Jaffe} \& {Perola}}{{Jaffe} \& {Perola}}{1973}]{1973jaffe}
{Jaffe} W.~J.,  {Perola} G.~C.,  1973, \aap, \href {https://ui.adsabs.harvard.edu/abs/1973A&A....26..423J} {26, 423}

\bibitem[\protect\citeauthoryear{{Jamrozy}, {Konar}, {Machalski}  \& {Saikia}}{{Jamrozy} et~al.}{2008}]{2008jamrozy}
{Jamrozy} M.,  {Konar} C.,  {Machalski} J.,   {Saikia} D.~J.,  2008, \mn@doi [\mnras] {10.1111/j.1365-2966.2007.12772.x}, \href {https://ui.adsabs.harvard.edu/abs/2008MNRAS.385.1286J} {385, 1286}

\bibitem[\protect\citeauthoryear{{Jetha}, {Hardcastle}, {Ponman}  \& {Sakelliou}}{{Jetha} et~al.}{2008}]{2008jetha}
{Jetha} N.~N.,  {Hardcastle} M.~J.,  {Ponman} T.~J.,   {Sakelliou} I.,  2008, \mn@doi [\mnras] {10.1111/j.1365-2966.2008.13959.x}, \href {https://ui.adsabs.harvard.edu/abs/2008MNRAS.391.1052J} {391, 1052}

\bibitem[\protect\citeauthoryear{{Jones} \& {Owen}}{{Jones} \& {Owen}}{1979}]{1979jones}
{Jones} T.~W.,  {Owen} F.~N.,  1979, \mn@doi [\apj] {10.1086/157561}, \href {https://ui.adsabs.harvard.edu/abs/1979ApJ...234..818J} {234, 818}

\bibitem[\protect\citeauthoryear{{Kale} \& {Ishwara-Chandra}}{{Kale} \& {Ishwara-Chandra}}{2021}]{2021kale}
{Kale} R.,  {Ishwara-Chandra} C.~H.,  2021, \mn@doi [Experimental Astronomy] {10.1007/s10686-020-09677-6}, \href {https://ui.adsabs.harvard.edu/abs/2021ExA....51...95K} {51, 95}

\bibitem[\protect\citeauthoryear{{Kim}, {Cassity}, {Bhatt}, {Fabbiano}, {Martinez Galarza}, {O'Sullivan}  \& {Rots}}{{Kim} et~al.}{2023}]{2023kim}
{Kim} D.-W.,  {Cassity} A.,  {Bhatt} B.,  {Fabbiano} G.,  {Martinez Galarza} J.~R.,  {O'Sullivan} E.,   {Rots} A.,  2023, \mn@doi [\apjs] {10.3847/1538-4365/ace4cc}, \href {https://ui.adsabs.harvard.edu/abs/2023ApJS..268...17K} {268, 17}

\bibitem[\protect\citeauthoryear{{Kirkpatrick} et~al.,}{{Kirkpatrick} et~al.}{2021}]{2021kirkpatrick}
{Kirkpatrick} C.~C.,  et~al., 2021, \mn@doi [\mnras] {10.1093/mnras/stab127}, \href {https://ui.adsabs.harvard.edu/abs/2021MNRAS.503.5763K} {503, 5763}

\bibitem[\protect\citeauthoryear{{Klein}, {Mack}, {Gregorini}  \& {Vigotti}}{{Klein} et~al.}{2003}]{2003klein}
{Klein} U.,  {Mack} K.~H.,  {Gregorini} L.,   {Vigotti} M.,  2003, \mn@doi [\aap] {10.1051/0004-6361:20030825}, \href {https://ui.adsabs.harvard.edu/abs/2003A&A...406..579K} {406, 579}

\bibitem[\protect\citeauthoryear{{Kolokythas}, {O'Sullivan}, {Giacintucci}, {Raychaudhury}, {Ishwara-Chandra}, {Worrall}  \& {Birkinshaw}}{{Kolokythas} et~al.}{2015}]{2015kolokythas}
{Kolokythas} K.,  {O'Sullivan} E.,  {Giacintucci} S.,  {Raychaudhury} S.,  {Ishwara-Chandra} C.~H.,  {Worrall} D.~M.,   {Birkinshaw} M.,  2015, \mn@doi [\mnras] {10.1093/mnras/stv665}, \href {https://ui.adsabs.harvard.edu/abs/2015MNRAS.450.1732K} {450, 1732}

\bibitem[\protect\citeauthoryear{{Kozie{\l}-Wierzbowska} \& {Stasi{\'n}ska}}{{Kozie{\l}-Wierzbowska} \& {Stasi{\'n}ska}}{2011}]{2011koziel}
{Kozie{\l}-Wierzbowska} D.,  {Stasi{\'n}ska} G.,  2011, \mn@doi [\mnras] {10.1111/j.1365-2966.2011.18346.x}, \href {https://ui.adsabs.harvard.edu/abs/2011MNRAS.415.1013K} {415, 1013}

\bibitem[\protect\citeauthoryear{{Leahy} \& {Williams}}{{Leahy} \& {Williams}}{1984}]{1984leahy}
{Leahy} J.~P.,  {Williams} A.~G.,  1984, \mn@doi [\mnras] {10.1093/mnras/210.4.929}, \href {https://ui.adsabs.harvard.edu/abs/1984MNRAS.210..929L} {210, 929}

\bibitem[\protect\citeauthoryear{{Ma} et~al.,}{{Ma} et~al.}{2019}]{2019mazhixian}
{Ma} Z.,  et~al., 2019, \mn@doi [\apjs] {10.3847/1538-4365/aaf9a2}, \href {https://ui.adsabs.harvard.edu/abs/2019ApJS..240...34M} {240, 34}

\bibitem[\protect\citeauthoryear{{McMullin}, {Waters}, {Schiebel}, {Young}  \& {Golap}}{{McMullin} et~al.}{2007}]{2007mcmullin}
{McMullin} J.~P.,  {Waters} B.,  {Schiebel} D.,  {Young} W.,   {Golap} K.,  2007, in {Shaw} R.~A.,  {Hill} F.,   {Bell} D.~J.,  eds,  Astronomical Society of the Pacific Conference Series Vol. 376, Astronomical Data Analysis Software and Systems XVI. p.~127

\bibitem[\protect\citeauthoryear{{Miley}}{{Miley}}{1980}]{1980miley}
{Miley} G.,  1980, \mn@doi [\araa] {10.1146/annurev.aa.18.090180.001121}, \href {https://ui.adsabs.harvard.edu/abs/1980ARA&A..18..165M} {18, 165}

\bibitem[\protect\citeauthoryear{{Miley}, {Perola}, {van der Kruit}  \& {van der Laan}}{{Miley} et~al.}{1972}]{1972miley}
{Miley} G.~K.,  {Perola} G.~C.,  {van der Kruit} P.~C.,   {van der Laan} H.,  1972, \mn@doi [\nat] {10.1038/237269a0}, \href {https://ui.adsabs.harvard.edu/abs/1972Natur.237..269M} {237, 269}

\bibitem[\protect\citeauthoryear{{Mingo} et~al.,}{{Mingo} et~al.}{2019}]{2019mingo}
{Mingo} B.,  et~al., 2019, \mn@doi [\mnras] {10.1093/mnras/stz1901}, \href {https://ui.adsabs.harvard.edu/abs/2019MNRAS.488.2701M} {488, 2701}

\bibitem[\protect\citeauthoryear{{Miraghaei} \& {Best}}{{Miraghaei} \& {Best}}{2017}]{2017miraghaei}
{Miraghaei} H.,  {Best} P.~N.,  2017, \mn@doi [\mnras] {10.1093/mnras/stx007}, \href {https://ui.adsabs.harvard.edu/abs/2017MNRAS.466.4346M} {466, 4346}

\bibitem[\protect\citeauthoryear{{Missaglia}, {Massaro}, {Capetti}, {Paolillo}, {Kraft}, {Baldi}  \& {Paggi}}{{Missaglia} et~al.}{2019}]{2019missaglia}
{Missaglia} V.,  {Massaro} F.,  {Capetti} A.,  {Paolillo} M.,  {Kraft} R.~P.,  {Baldi} R.~D.,   {Paggi} A.,  2019, \mn@doi [\aap] {10.1051/0004-6361/201935058}, \href {https://ui.adsabs.harvard.edu/abs/2019A&A...626A...8M} {626, A8}

\bibitem[\protect\citeauthoryear{{Morris}, {Wilcots}, {Hooper}  \& {Heinz}}{{Morris} et~al.}{2022}]{2022morris}
{Morris} M.~E.,  {Wilcots} E.,  {Hooper} E.,   {Heinz} S.,  2022, \mn@doi [\aj] {10.3847/1538-3881/ac66db}, \href {https://ui.adsabs.harvard.edu/abs/2022AJ....163..280M} {163, 280}

\bibitem[\protect\citeauthoryear{{Murgia}, {Fanti}, {Fanti}, {Gregorini}, {Klein}, {Mack}  \& {Vigotti}}{{Murgia} et~al.}{1999}]{1999murgia}
{Murgia} M.,  {Fanti} C.,  {Fanti} R.,  {Gregorini} L.,  {Klein} U.,  {Mack} K.~H.,   {Vigotti} M.,  1999, \aap, \href {https://ui.adsabs.harvard.edu/abs/1999A&A...345..769M} {345, 769}

\bibitem[\protect\citeauthoryear{{Murgia} et~al.,}{{Murgia} et~al.}{2011}]{2011murgia}
{Murgia} M.,  et~al., 2011, \mn@doi [\aap] {10.1051/0004-6361/201015302}, \href {https://ui.adsabs.harvard.edu/abs/2011A&A...526A.148M} {526, A148}

\bibitem[\protect\citeauthoryear{{O'Brien}, {Norris}, {Tothill}  \& {Filipovi{\'c}}}{{O'Brien} et~al.}{2018}]{2018obrien}
{O'Brien} A.~N.,  {Norris} R.~P.,  {Tothill} N. F.~H.,   {Filipovi{\'c}} M.~D.,  2018, \mn@doi [\mnras] {10.1093/mnras/sty2642}, \href {https://ui.adsabs.harvard.edu/abs/2018MNRAS.481.5247O} {481, 5247}

\bibitem[\protect\citeauthoryear{{O'Dea} \& {Baum}}{{O'Dea} \& {Baum}}{2023}]{2023odea}
{O'Dea} C.~P.,  {Baum} S.~A.,  2023, \mn@doi [Galaxies] {10.3390/galaxies11030067}, \href {https://ui.adsabs.harvard.edu/abs/2023Galax..11...67O} {11, 67}

\bibitem[\protect\citeauthoryear{{O'Donoghue}, {Eilek}  \& {Owen}}{{O'Donoghue} et~al.}{1993}]{1993odonoghue}
{O'Donoghue} A.~A.,  {Eilek} J.~A.,   {Owen} F.~N.,  1993, \mn@doi [\apj] {10.1086/172600}, \href {https://ui.adsabs.harvard.edu/abs/1993ApJ...408..428O} {408, 428}

\bibitem[\protect\citeauthoryear{{Owen} \& {Rudnick}}{{Owen} \& {Rudnick}}{1976}]{1976owen}
{Owen} F.~N.,  {Rudnick} L.,  1976, \mn@doi [\apjl] {10.1086/182077}, \href {https://ui.adsabs.harvard.edu/abs/1976ApJ...205L...1O} {205, L1}

\bibitem[\protect\citeauthoryear{{Pandge}, {Kale}, {Dabhade}, {Mahato}  \& {Raychaudhury}}{{Pandge} et~al.}{2022}]{2022pandge}
{Pandge} M.~B.,  {Kale} R.,  {Dabhade} P.,  {Mahato} M.,   {Raychaudhury} S.,  2022, \mn@doi [\mnras] {10.1093/mnras/stab2945}, \href {https://ui.adsabs.harvard.edu/abs/2022MNRAS.509.1837P} {509, 1837}

\bibitem[\protect\citeauthoryear{{Perley} \& {Butler}}{{Perley} \& {Butler}}{2017}]{2017perley}
{Perley} R.~A.,  {Butler} B.~J.,  2017, \mn@doi [\apjs] {10.3847/1538-4365/aa6df9}, \href {https://ui.adsabs.harvard.edu/abs/2017ApJS..230....7P} {230, 7}

\bibitem[\protect\citeauthoryear{{Pinkney}, {Burns}  \& {Hill}}{{Pinkney} et~al.}{1994}]{1994pinkney}
{Pinkney} J.,  {Burns} J.~O.,   {Hill} J.~M.,  1994, \mn@doi [\aj] {10.1086/117216}, \href {https://ui.adsabs.harvard.edu/abs/1994AJ....108.2031P} {108, 2031}

\bibitem[\protect\citeauthoryear{{Planck Collaboration} et~al.,}{{Planck Collaboration} et~al.}{2016}]{2016planck}
{Planck Collaboration} et~al., 2016, \mn@doi [\aap] {10.1051/0004-6361/201525830}, \href {https://ui.adsabs.harvard.edu/abs/2016A&A...594A..13P} {594, A13}

\bibitem[\protect\citeauthoryear{{Raj}, {Ishwara-Chandra}, {Sudheesh}, {Biju}  \& {Jacob}}{{Raj} et~al.}{2025}]{2025raj}
{Raj} A.,  {Ishwara-Chandra} C.~H.,  {Sudheesh} T.~P.,  {Biju} K.~G.,   {Jacob} J.,  2025, \mn@doi [Journal of Astrophysics and Astronomy] {10.1007/s12036-024-10035-7}, \href {https://ui.adsabs.harvard.edu/abs/2025JApA...46....7R} {46, 7}

\bibitem[\protect\citeauthoryear{{Reddy} et~al.,}{{Reddy} et~al.}{2017}]{2017reddy}
{Reddy} S.~H.,  et~al., 2017, \mn@doi [Journal of Astronomical Instrumentation] {10.1142/S2251171716410117}, \href {https://ui.adsabs.harvard.edu/abs/2017JAI.....641011R} {6, 1641011}

\bibitem[\protect\citeauthoryear{{Riley}}{{Riley}}{1972}]{1972riley}
{Riley} J.~M.,  1972, \mn@doi [\mnras] {10.1093/mnras/157.4.349}, \href {https://ui.adsabs.harvard.edu/abs/1972MNRAS.157..349R} {157, 349}

\bibitem[\protect\citeauthoryear{{Robertson}}{{Robertson}}{1984}]{1984robertson}
{Robertson} J.~G.,  1984, \aap, \href {https://ui.adsabs.harvard.edu/abs/1984A&A...138...41R} {138, 41}

\bibitem[\protect\citeauthoryear{{Robitaille} \& {Bressert}}{{Robitaille} \& {Bressert}}{2012}]{2012robitaille}
{Robitaille} T.,  {Bressert} E.,  2012, {APLpy: Astronomical Plotting Library in Python}, Astrophysics Source Code Library (\mn@eprint {ascl} {1208.017})

\bibitem[\protect\citeauthoryear{{Ryle} \& {Windram}}{{Ryle} \& {Windram}}{1968}]{1968ryle}
{Ryle} M.,  {Windram} M.~D.,  1968, \mn@doi [\mnras] {10.1093/mnras/138.1.1}, \href {https://ui.adsabs.harvard.edu/abs/1968MNRAS.138....1R} {138, 1}

\bibitem[\protect\citeauthoryear{{Saikia} \& {Jamrozy}}{{Saikia} \& {Jamrozy}}{2009}]{2009saikia}
{Saikia} D.~J.,  {Jamrozy} M.,  2009, \mn@doi [Bulletin of the Astronomical Society of India] {10.48550/arXiv.1002.1841}, \href {https://ui.adsabs.harvard.edu/abs/2009BASI...37...63S} {37, 63}

\bibitem[\protect\citeauthoryear{{Sakelliou}, {Merrifield}  \& {McHardy}}{{Sakelliou} et~al.}{1996}]{1996sakelliou}
{Sakelliou} I.,  {Merrifield} M.~R.,   {McHardy} I.~M.,  1996, \mn@doi [\mnras] {10.1093/mnras/283.2.673}, \href {https://ui.adsabs.harvard.edu/abs/1996MNRAS.283..673S} {283, 673}

\bibitem[\protect\citeauthoryear{{Santra} et~al.,}{{Santra} et~al.}{2024}]{2024santra}
{Santra} R.,  et~al., 2024, \mn@doi [\apj] {10.3847/1538-4357/ad1190}, \href {https://ui.adsabs.harvard.edu/abs/2024ApJ...962...40S} {962, 40}

\bibitem[\protect\citeauthoryear{{Scheuer}}{{Scheuer}}{1974}]{1974scheuer}
{Scheuer} P.~A.~G.,  1974, \mn@doi [\mnras] {10.1093/mnras/166.3.513}, \href {https://ui.adsabs.harvard.edu/abs/1974MNRAS.166..513S} {166, 513}

\bibitem[\protect\citeauthoryear{{Shimwell} et~al.,}{{Shimwell} et~al.}{2019}]{2019shimwell}
{Shimwell} T.~W.,  et~al., 2019, \mn@doi [\aap] {10.1051/0004-6361/201833559}, \href {https://ui.adsabs.harvard.edu/abs/2019A&A...622A...1S} {622, A1}

\bibitem[\protect\citeauthoryear{{Shimwell} et~al.,}{{Shimwell} et~al.}{2022}]{2022shimwell}
{Shimwell} T.~W.,  et~al., 2022, \mn@doi [\aap] {10.1051/0004-6361/202142484}, \href {https://ui.adsabs.harvard.edu/abs/2022A&A...659A...1S} {659, A1}

\bibitem[\protect\citeauthoryear{{Tempel}, {Tuvikene}, {Kipper}  \& {Libeskind}}{{Tempel} et~al.}{2017}]{2017tempel}
{Tempel} E.,  {Tuvikene} T.,  {Kipper} R.,   {Libeskind} N.~I.,  2017, \mn@doi [\aap] {10.1051/0004-6361/201730499}, \href {https://ui.adsabs.harvard.edu/abs/2017A&A...602A.100T} {602, A100}

\bibitem[\protect\citeauthoryear{{Turner}, {Shabala}  \& {Krause}}{{Turner} et~al.}{2018}]{2018turner}
{Turner} R.~J.,  {Shabala} S.~S.,   {Krause} M. G.~H.,  2018, \mn@doi [\mnras] {10.1093/mnras/stx2947}, \href {https://ui.adsabs.harvard.edu/abs/2018MNRAS.474.3361T} {474, 3361}

\bibitem[\protect\citeauthoryear{{Venkatesan}, {Batuski}, {Hanisch}  \& {Burns}}{{Venkatesan} et~al.}{1994}]{1994venkatesan}
{Venkatesan} T.~C.~A.,  {Batuski} D.~J.,  {Hanisch} R.~J.,   {Burns} J.~O.,  1994, \mn@doi [\apj] {10.1086/174881}, \href {https://ui.adsabs.harvard.edu/abs/1994ApJ...436...67V} {436, 67}

\bibitem[\protect\citeauthoryear{{Wen}, {Han}  \& {Liu}}{{Wen} et~al.}{2012}]{2012wen}
{Wen} Z.~L.,  {Han} J.~L.,   {Liu} F.~S.,  2012, \mn@doi [\apjs] {10.1088/0067-0049/199/2/34}, \href {https://ui.adsabs.harvard.edu/abs/2012ApJS..199...34W} {199, 34}

\bibitem[\protect\citeauthoryear{{Wing} \& {Blanton}}{{Wing} \& {Blanton}}{2011}]{2011wing}
{Wing} J.~D.,  {Blanton} E.~L.,  2011, \mn@doi [\aj] {10.1088/0004-6256/141/3/88}, \href {https://ui.adsabs.harvard.edu/abs/2011AJ....141...88W} {141, 88}

\bibitem[\protect\citeauthoryear{{Worrall}, {Birkinshaw}  \& {Cameron}}{{Worrall} et~al.}{1995}]{1995worrall}
{Worrall} D.~M.,  {Birkinshaw} M.,   {Cameron} R.~A.,  1995, \mn@doi [\apj] {10.1086/176035}, \href {https://ui.adsabs.harvard.edu/abs/1995ApJ...449...93W} {449, 93}

\bibitem[\protect\citeauthoryear{{Zhao}, {Arag{\'o}n-Salamanca}  \& {Conselice}}{{Zhao} et~al.}{2015}]{2015zhao}
{Zhao} D.,  {Arag{\'o}n-Salamanca} A.,   {Conselice} C.~J.,  2015, \mn@doi [\mnras] {10.1093/mnras/stv190}, \href {https://ui.adsabs.harvard.edu/abs/2015MNRAS.448.2530Z} {448, 2530}

\bibitem[\protect\citeauthoryear{{de Gasperin} et~al.,}{{de Gasperin} et~al.}{2017}]{2017degasperin}
{de Gasperin} F.,  et~al., 2017, \mn@doi [Science Advances] {10.1126/sciadv.1701634}, \href {https://ui.adsabs.harvard.edu/abs/2017SciA....3E1634D} {3, e1701634}

\makeatother
\end{thebibliography}




\appendix

\section{Regions used for estimating flux density}
In Figures \ref{fig:region-for-flux-density}, \ref{fig:nvss-tgss-radiomaps}, and \ref{fig:lobe-fluxdensity-regions}, we show the regions used to estimate the flux densities from the radio maps. Fig.~\ref{fig:region-for-flux-density} (top) shows the uGMRT Band 3 radio map with the region overlaid, while Fig.~\ref{fig:region-for-flux-density} (bottom) shows the uGMRT Band 4 map with the same region overlaid. Fig.~\ref{fig:nvss-tgss-radiomaps} (top) shows the TGSS radio map, and Fig.~\ref{fig:nvss-tgss-radiomaps} (bottom) shows the NVSS map, both with the estimation region overlaid. In all the radio maps shown in Figs.~\ref{fig:region-for-flux-density} and \ref{fig:nvss-tgss-radiomaps}, the same region is used for estimating flux density. Fig.~\ref{fig:lobe-fluxdensity-regions} shows the uGMRT Band 3 map with two regions overlaid, which were used to estimate the lobe flux densities. The same regions were also used for lobe flux density estimation from the uGMRT Band 4, TGSS, and NVSS maps.

\begin{center}
    \includegraphics[width=\columnwidth]{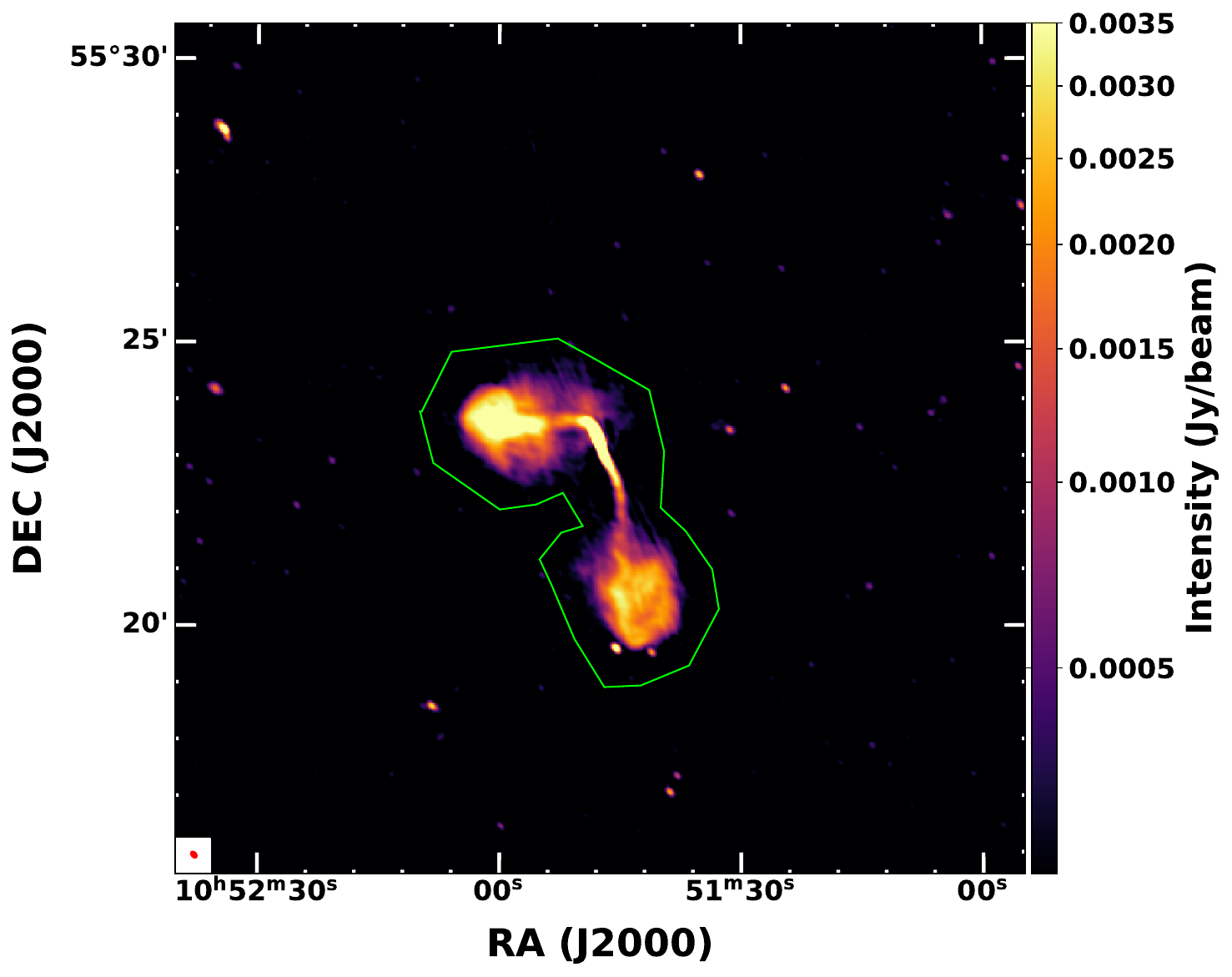}\\
    \includegraphics[width=\columnwidth]{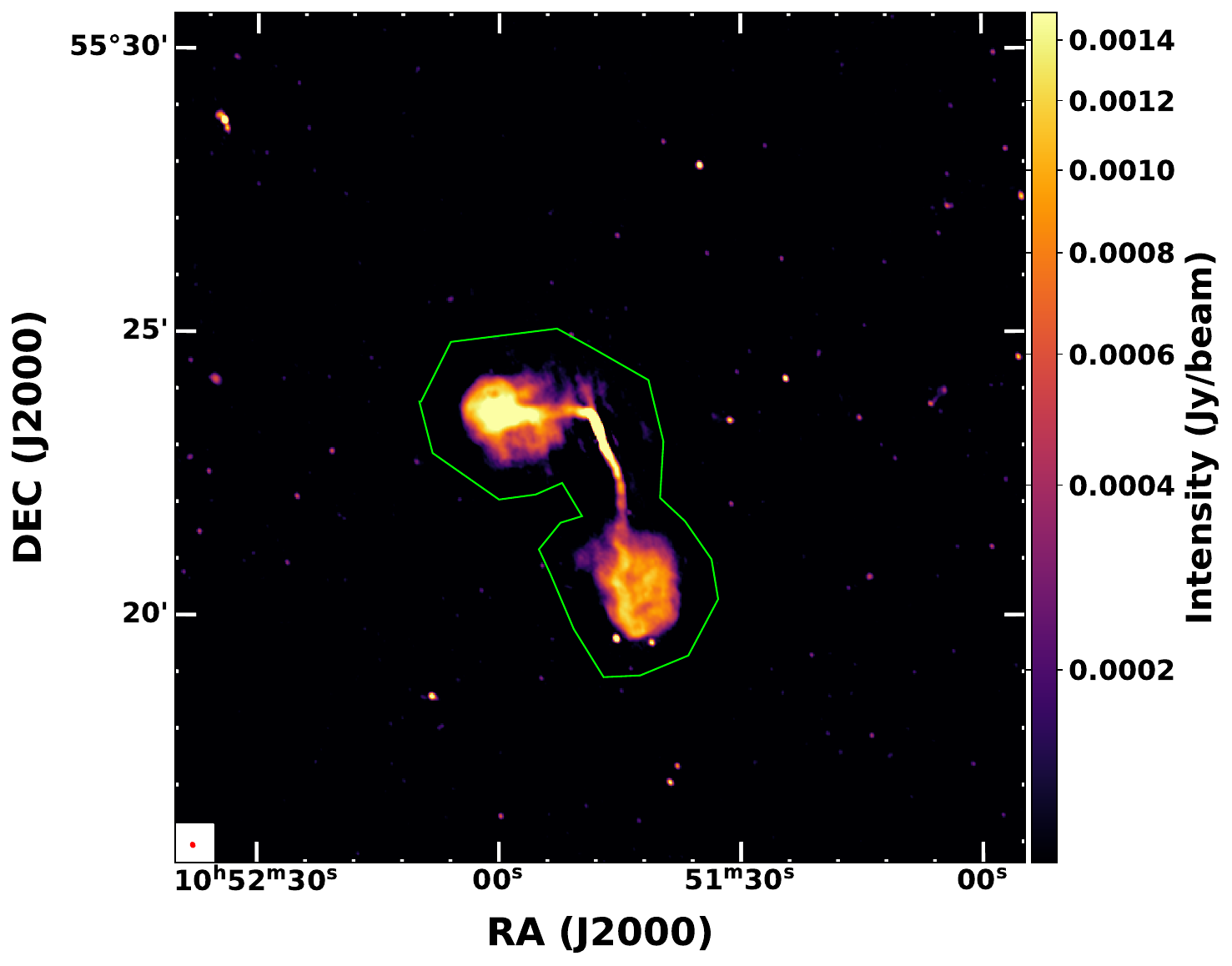}
    \captionof{figure}{Top panel: uMGRT Band 3 radio map. Bottom panel: uMGRT Band 4 radio map. The green contours overlaid over the images shows the regions used for extracting the source flux densities from the radio maps.}
    \label{fig:region-for-flux-density}
\end{center}

\begin{center}
    \includegraphics[width=\columnwidth]{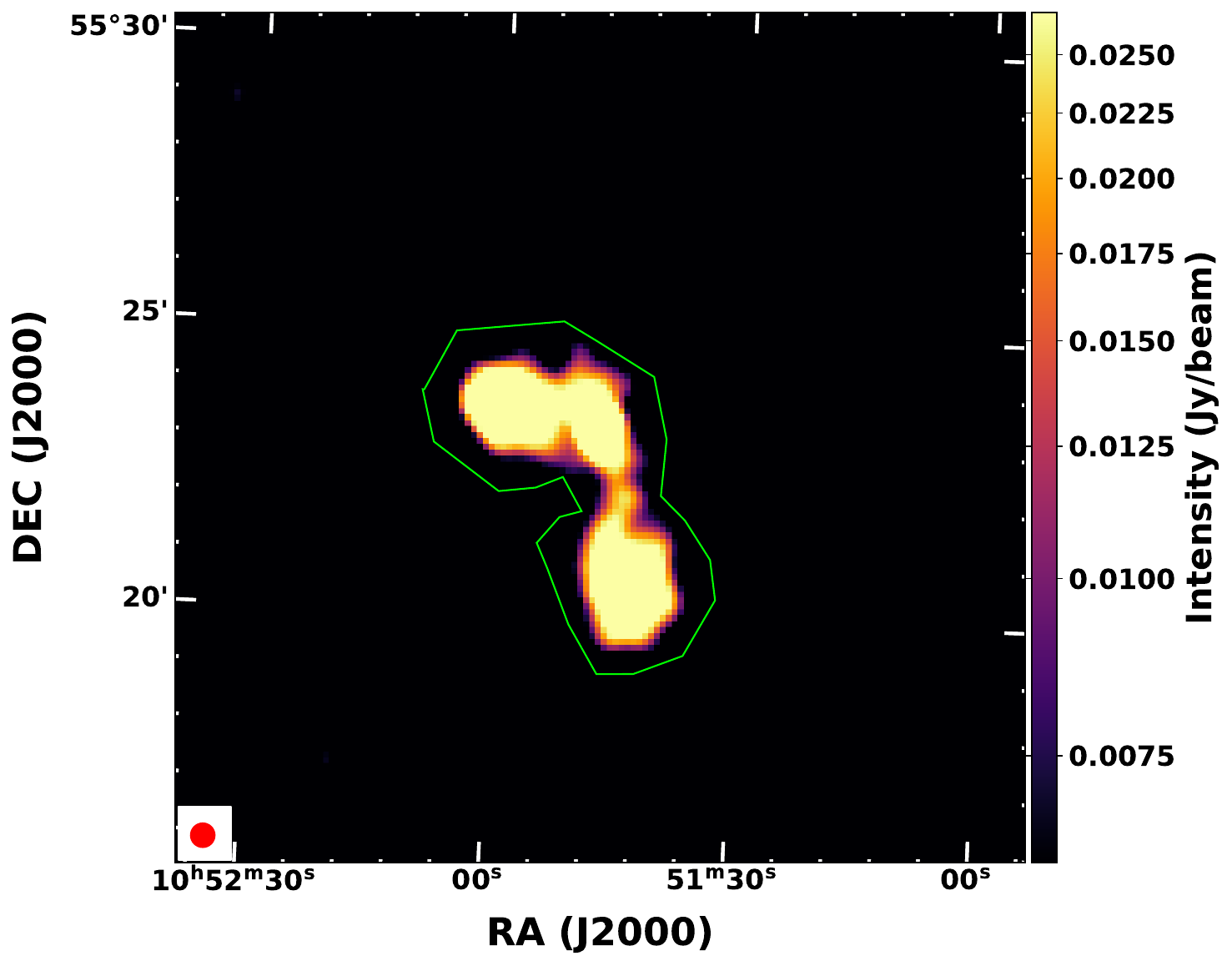}\\
    \includegraphics[width=\columnwidth]{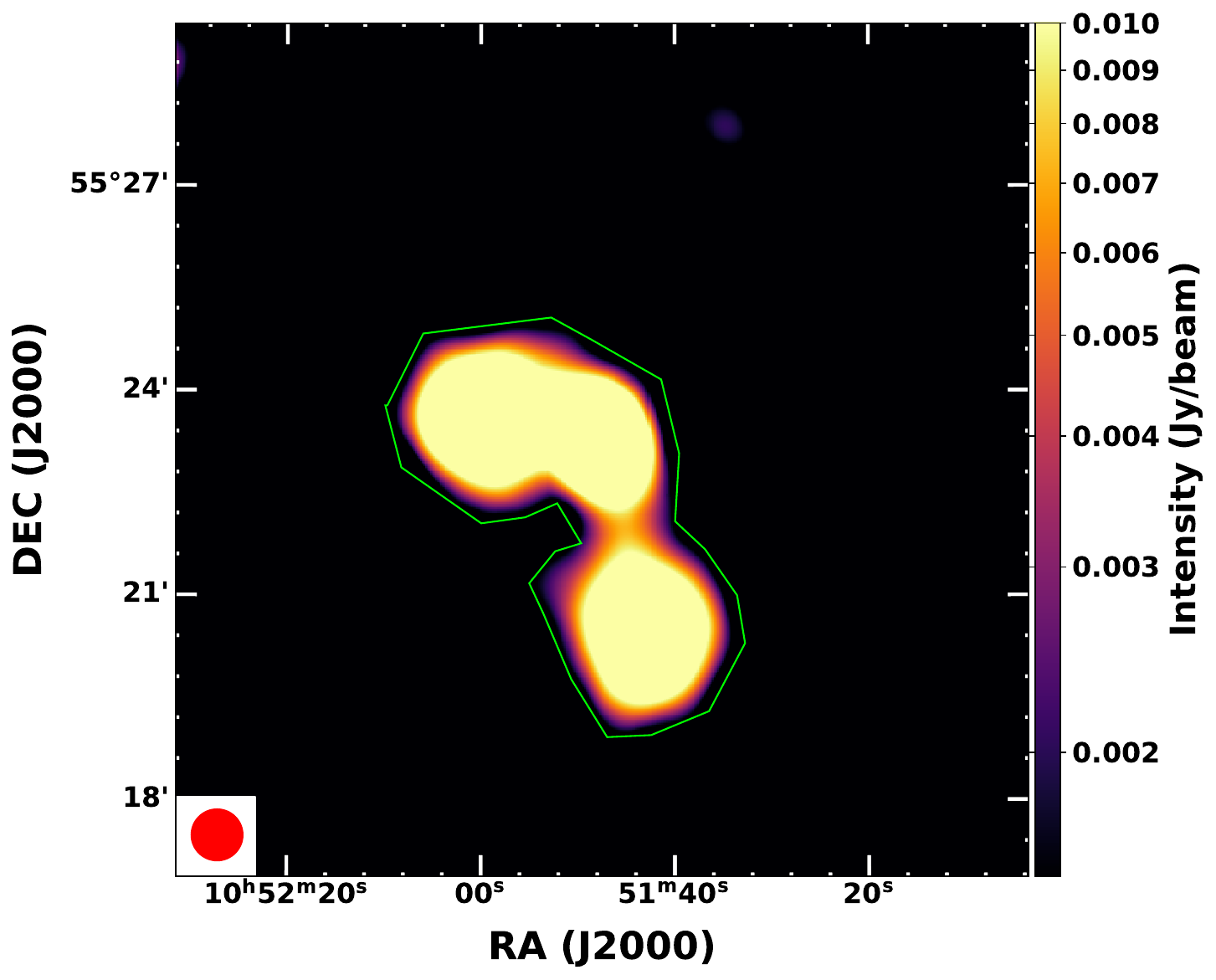}
    \captionof{figure}{Top panel: TGSS radio map of J1051+5523. Bottom panel: NVSS radio map of J1051+5523. The green contours overlaid over the images shows the regions used for extracting the source flux densities from the radio maps. The radio beams are provided at the bottom left corner of the images. The radio parameters measured from the images are provided in Table~\ref{tab:observed-spectrum}.}
    \label{fig:nvss-tgss-radiomaps}
\end{center}

\begin{center}
    \includegraphics[width=\columnwidth]{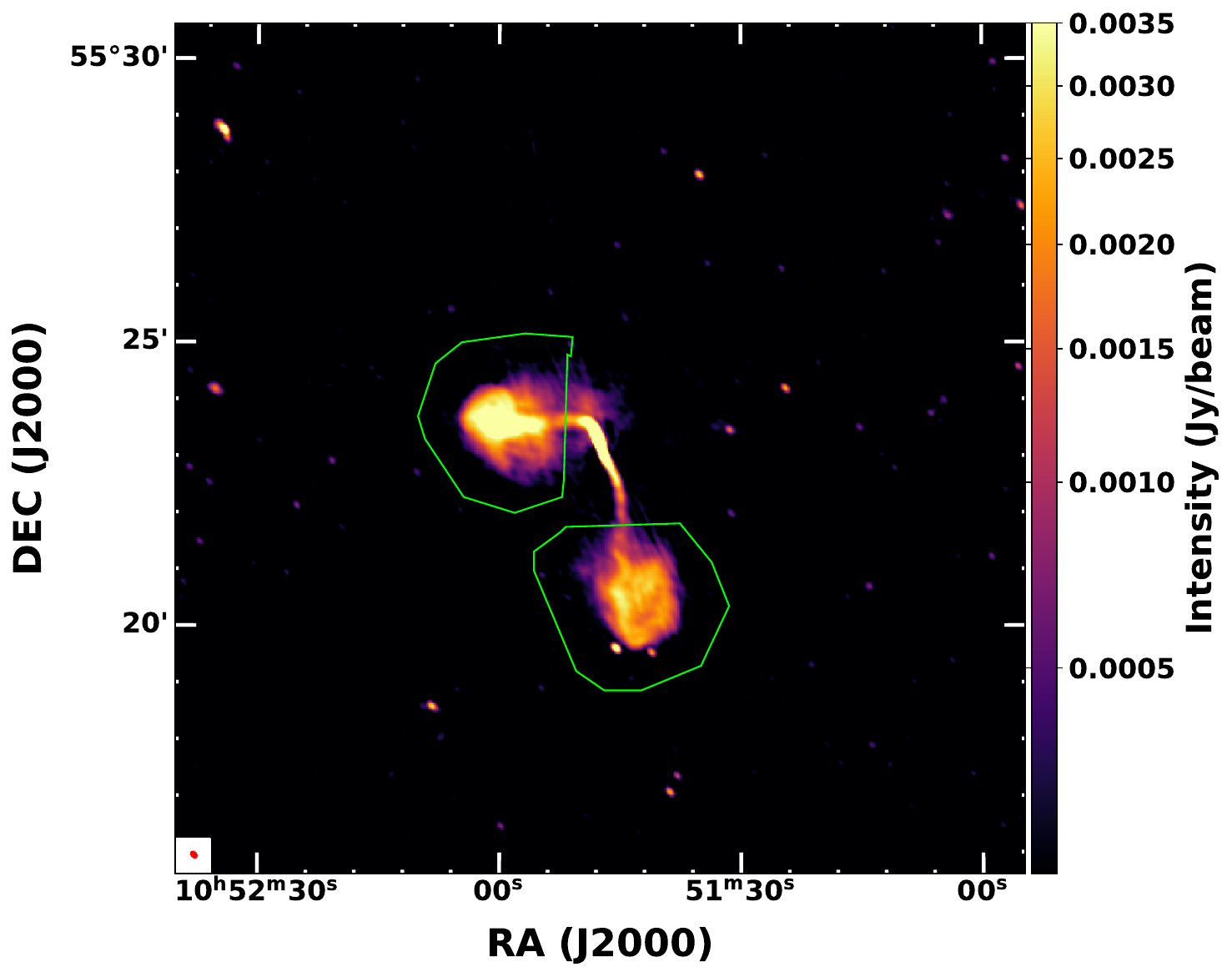}
    \captionof{figure}{uGMRT Band 3 radio map of J1051+5523. The green contours overlaid over the image shows the regions used for extracting the lobe flux densities from the radio map. The radio beam is provided at the bottom left corner of the image. The same regions are used for extracting the lobe flux densities from uGMRT Band 4, TGSS, and NVSS radio maps. The radio parameters measured from the images are provided in Table~\ref{tab:lobe-flux-density}.}
    \label{fig:lobe-fluxdensity-regions}
\end{center}


\bsp	
\label{lastpage}
\end{document}